\documentclass{article}

\usepackage[main, final]{neurips_2026}

\usepackage[utf8]{inputenc} 
\usepackage[T1]{fontenc}    
\usepackage{hyperref}       
\usepackage{url}            
\usepackage{booktabs}       
\usepackage{amsfonts}       
\usepackage{nicefrac}       
\usepackage{microtype}      
\usepackage{xcolor}         

\usepackage{array}
\usepackage{tabularx}
\usepackage{caption}

\usepackage{graphicx}
\usepackage{subcaption}
\usepackage{epstopdf}
\usepackage{etoc}
\usepackage{listings}
\usepackage{amsmath}
\usepackage[most]{tcolorbox}
\usepackage{tcolorbox}
\usepackage{listings}
\usepackage{xcolor}

\definecolor{promptblue}{RGB}{68,142,201}
\definecolor{promptbg}{RGB}{234,241,248}
\newtcblisting{promptbox}[2][]{
    enhanced,
    breakable,
    colback=promptbg,
    colframe=promptblue,
    colbacktitle=promptblue,
    coltitle=white,
    title={#2},
    boxrule=0.8pt,
    arc=1.2mm,
    left=2mm,
    right=2mm,
    top=1mm,
    bottom=1mm,
    listing only,
    listing options={
        basicstyle=\small\ttfamily,
        breaklines=true,
        columns=fullflexible,
        keepspaces=true,
        showstringspaces=false
    },
    #1
}

\title{Token Inflation: How Dishonest Providers Can Overcharge for Large Language Model Usage}

\author{
  \textbf{Shahinul Hoque}$^{1}$,
  \textbf{Jinghuai Zhang}$^{2}$,
  \textbf{Jinyuan Sun}$^{1}$,
  \textbf{Fnu Suya}$^{1}$ \\
  $^{1}$University of Tennessee, Knoxville,
  $^{2}$University of California, Los Angeles \\
  \texttt{shoque@vols.utk.edu,
  jinghuai1998@g.ucla.edu,
  jysun@utk.edu,
  fsuya@utk.edu}
}

\begin{document}

\maketitle

\begin{abstract}

Per-token billing is now the standard pricing model for commercial large language models (LLMs), so the honesty of reported token counts directly affects what users pay. We show that this kind of billing is hard to audit by design: providers hide the model, the tokenizer, and the execution to protect their IP, mitigate jailbreaks, and preserve user privacy, which means an auditor can only inspect proofs the provider supplies. The audit therefore reduces to a consistency check on the provider's own reports. We call this a \emph{trust paradox}: every audit must trust some artifact, but current frameworks trust exactly the ones a provider has the strongest reason to manipulate. We study three recent token auditing frameworks and show that a provider with ordinary commercial capabilities can systematically inflate billed token counts. In the most permissive setting, hidden reasoning usage can be inflated by 1{,}469\% on average without detection. At current frontier reasoning prices, that turns a \$100 honest bill into roughly a \$1{,}569 bill on the same query. Even when the user can see the full reasoning string, tokenization ambiguity alone still allows 50.85\% over-reporting below the detection threshold. These results suggest the problem is not in any specific auditor but in any audit whose evidence comes from the audited party. Restoring honest billing will require verification that ties reported token counts to evidence the provider does not control, such as trusted execution attestation, cryptographic proofs of inference, or third-party re-execution.

\end{abstract}

\section{Introduction}

Per-token billing has become the dominant pricing model for commercial large language models (LLMs), with frontier reasoning models charging tens of dollars per million tokens and processing large volumes of billed traffic across consumer and agentic deployments~\cite{bao2026nvidia_tokens,guo2025deepseek}. Even small inflation of reported token counts therefore translates into substantial overcharges at scale, making the integrity of reported token counts an economically consequential question.

The same services that bill per token also have strong reasons to keep their internal computation hidden. Exposing the full reasoning trace would help competitors train similar models from it~\cite{guo2025deepseek,carlini2024stealing}, make it easier to attack safety mechanisms through chain-of-thought hijacking~\cite{kuo2025h}, and risk leaking private user information from intermediate steps~\cite{tamber2025can}. Today's commercial reasoning APIs reflect this. OpenAI's o1 and o3 models return only a short summary of their reasoning while still billing the user for the full hidden trace. Google's Gemini 2.5 thinking models likewise return summaries instead of raw thoughts. Even Anthropic's extended thinking interface only partially exposes the reasoning content. The user typically sees a short final answer while the bill includes many times more reasoning tokens that the user never gets to see, so any inflation in those hidden tokens is invisible by design.

Auditing in this setting has a structural disadvantage. The opacity that creates the need for auditing is the same opacity that blocks it. Without access to the model, tokenizer, or execution, current auditors must rely on indirect evidence rather than verifying actual token usage. This leads to a \emph{proxy gap}: satisfying an auditor's check does not necessarily establish that the reported tokens were actually generated. The problem is compounded by a \emph{trust paradox}: the evidence used by the auditor is often supplied or influenced by the very provider whose billing behavior it is meant to verify.

We demonstrate the consequences of these weaknesses across three recent state-of-the-art auditing frameworks, CoIn~\cite{sun2025coin}, PALACE~\cite{wang2025predictive}, and a martingale-based statistical auditor~\cite{velasco2025auditing}. In our evaluation, simple and inexpensive strategies inflate reported token counts by up to 1{,}469\% on average in the CoIn setting, by 247\% in the PALACE setting where the auditor trusts data supplied by the provider, and by 50.85\% even when the user can see the full reasoning string. At current output prices for frontier reasoning models, the CoIn case alone turns a \$100 honest reasoning bill into roughly a \$1{,}569 bill on the same query. These results demonstrate that the proxy gap can be exploited across different auditing designs, rather than being an isolated weakness of a particular framework.

\paragraph{Contributions.}

Our main contribution is a structural analysis of LLM token-count auditing, identifying a common \emph{proxy gap} between what auditors intend to verify and what their evidence can establish. Through systematic adversarial evaluations of three recent frameworks, we demonstrate how dishonest providers can exploit these gaps to inflate reported token counts while evading detection. First, CoIn can be evaded by reusing or fabricating plausible reasoning blocks, achieving 1{,}469\% average token-count inflation without detection (Section~\ref{sec:coin}). Second, PALACE can be manipulated through answer rewriting, trigger tokens, and provider-supplied training-data poisoning to inflate predicted reasoning-token counts (Section~\ref{sec:palace}). Third, a martingale-based statistical auditor can be evaded through strategic over- and under-reporting, allowing 50.85\% net token-count inflation even when the full reasoning string is visible (Section~\ref{sec:stat}). Finally, we evaluate practical defenses that narrow these gaps and show that provider-independent verification ultimately requires binding token counts to trusted execution (Section~\ref{sec:defense}). 

Together, these results show that greater transparency alone cannot guarantee accurate token-count auditing when the evidence remains under provider control. Stronger guarantees require binding reported token counts to independently verifiable evidence of actual inference computation.


\section{Threat Model and Related Work}
\label{sec:background}
We first introduce our threat model (Section \ref{sec:threat-model}) and then discuss the related work (Section \ref{sec:related work}).
\subsection{Threat Model}\label{sec:threat-model}

\paragraph{Attack settings.}
We consider an opaque LLM service where a user submits a prompt $P$, receives a final answer $A$, and is billed according to a provider-reported token count $m$. Internally, the provider may generate a reasoning trace $R$ that is hidden from the user but included in usage accounting. Thus, the provider observes $(P,R,A)$, while the user typically only observes $(P,A,m)$, and in some deployments $R$ as text. Depending on the service, $m$ may include prompt tokens, visible output tokens, hidden reasoning tokens, or other provider-side tokens.

Token count is a convenient billing abstraction because prompts, reasoning traces, retrieved context, tool calls, and final answers are all token sequences, and major APIs use per-token pricing~\cite{bao2026nvidia_tokens,anthropic_token_counting}. However, current commercial deployments do not let users directly verify the underlying tokenization or hidden reasoning. Even when $R$ is exposed as text, tokenization ambiguity~\cite{sennrich2016neural,kudo2018sentencepiece} can prevent a unique count, since distinct token sequences may decode to the same string. These verification challenges motivate the token-count auditing frameworks examined in this paper.

\paragraph{Attacker knowledge and capability.} We assume a financially motivated provider that may over-report usage. The provider operates the service end to end: it runs the model, tokenizes inputs and outputs, generates any hidden reasoning trace, formats the answer, and reports the bill. When the auditor is trained on provider-supplied data, as in PALACE (Section~\ref{sec:palace}), that data is also under provider control. In all settings, any provider-generated artifact used by an auditor, such as traces, embeddings, answers, auxiliary labels, or reported token counts, is part of the attack surface unless it is independently verifiable. Each framework section (section \ref{sec:coin}-\ref{sec:stat}) gives the framework-specific instantiation of these controls.

\begin{table}[ht]

\centering
\caption{Proxy Gap Across Token-Count Auditing Frameworks.}
\label{tab:proxy-gap}
\vspace{5pt}
\small
\setlength{\tabcolsep}{5pt}
\renewcommand{\arraystretch}{1.12}
\begin{tabularx}{\textwidth}{
@{}
>{\raggedright\arraybackslash}p{1.35cm}
>{\hsize=0.85\hsize\raggedright\arraybackslash}X
>{\hsize=0.85\hsize\raggedright\arraybackslash}X
>{\hsize=1.30\hsize\raggedright\arraybackslash}X
@{}}
\toprule
\textbf{Framework} &
\textbf{Proxy checked} &
\textbf{Provider control} &
\textbf{Proxy gap} \\
\midrule

CoIn &
Semantic validity of reasoning blocks &
Submitted reasoning trace &
Semantic validity does not prove genuine reasoning \\

PALACE &
Predicted hidden reasoning length &
Answer and auxiliary data &
Predicted length does not prove actual token count \\

Statistical auditor &
Aggregate consistency of reported counts &
Reported token counts &
Aggregate consistency does not prove per-request correctness \\

\bottomrule
\end{tabularx}
\end{table}

\paragraph{Adversarial objective and the proxy gap.}
A dishonest provider aims to report more tokens than were actually generated while still passing the audit. As Table~\ref{tab:proxy-gap} shows, the three existing frameworks rely on different proxies for the token quantity they intend to verify. Their vulnerabilities nevertheless share two \emph{conditions}. First, the token counting proxy can be satisfied without verifying actual token usage, creating a \emph{proxy gap}. Second, the provider controls the evidence evaluated by the auditor and can often evaluate the audit before reporting. The first condition creates room for dishonest reports, while the second makes that room inexpensive to exploit. Sections~\ref{sec:coin}--\ref{sec:stat} therefore follow a common structure: we identify the proxy and its gap, demonstrate a low-cost attack, consider a natural defense, and evaluate an adaptive attack showing that the defense narrows the gap without closing it.

\subsection{Related Work}\label{sec:related work}

Prior work on auditing opaque LLM services has primarily focused on two forms of provider-side dishonesty, first, token-count inflation in pay-per-use billing, and second, model substitution. In model substitution, a provider serves a cheaper model than advertised, potentially compromising output quality~\cite{cai2025you}. Existing approaches detect such substitutions through rank-based uniformity tests~\cite{zhu2025auditing} or secret-based verification~\cite{sun2024svip}.

Token-count inflation presents a different challenge where a provider can overcharge users without noticeably changing the returned output. Recent work highlights the risks of unverifiable token billing in opaque LLM services~\cite{sun2025invisible,velasco2025your}. CoIn~\cite{sun2025coin} and PALACE~\cite{wang2025predictive} audit hidden reasoning-token counts without direct access to the reasoning trace, while \citet{velasco2025auditing} study token-count auditing when the generated reasoning trace is visible but its tokenization remains ambiguous.

Our work builds on these token-count auditing frameworks and examines their security against dishonest providers. Rather than proposing another auditor, we identify common weaknesses in the evidence these frameworks rely on and demonstrate how providers can exploit them to inflate reported token counts while evading detection.


\section{Detecting Token Count Inflation via Semantic Similarity}
\label{sec:coin}
We first describe CoIn and its proxy gap (Section~\ref{sec:coin-overview}), then show a naive attack, a natural defense, and an adaptive attack that restores inflation (Section~\ref{sec:coin-simple-attack}), and finally other variants that exploit the same gap (Section~\ref{sec:coin-other-attacks}).

\subsection{Auditing Framework Overview and Proxy Gap}\label{sec:coin-overview}
CoIn~\cite{sun2025coin} audits services that return only the final answer while hiding the reasoning trace, with a trusted third-party auditor verifying the billed hidden tokens. The provider commits the trace as fixed-length token blocks in a Merkle hash tree (\emph{token-quantity verification}), and the auditor accepts or rejects it by scoring sampled blocks with two matching head scorers, token-to-block (T2B) and block-to-answer (B2A), followed by a rule-based or learned verifier (\emph{semantic-validity verification}).

\paragraph{Proxy gap.}
CoIn aims to verify that committed blocks reflect the model’s actual reasoning, but its matching heads assess only semantic similarity. \emph{Condition (i)}, the proxy gap, holds because similarity measures closeness in meaning, not whether a block was produced during reasoning~\cite{steck2024cosine}. A duplicate block can score identical to its original, and similarly a block generated from a different model's trace from the same prompt may score similar to the honest reasoning trace. \emph{Condition (ii)}, provider control, follows directly from the protocol. The Merkle commitment binds the provider to a chosen trace but does not verify that the trace was actually produced during inference. Because the provider selects the trace and can compute the auditor's embeddings before committing, it can locally evaluate candidate traces and commit only to one that passes. Together, these conditions allow the provider to choose audited traces from a set that extends beyond actual reasoning produced during inference. The audit therefore verifies that a passing trace was found, not that the model actually generated it during inference.

\subsection{Block-Reuse Attack and Adaptive Evasion}
\label{sec:coin-simple-attack}

We use CoIn's official models and framework with the Glaive reasoning-v1-20m dataset~\cite{glaive2025reasoningv120m}, one of the public datasets used in the original CoIn paper. Following the original setup, each reasoning trace is partitioned into fixed-length 256-token blocks using CoIn's block-generation procedure. About 97\% of evaluated samples contain between two and seven reasoning blocks, and we restrict our analysis to this range to avoid sparsely populated block-count regions. Inflation is reported as the percentage increase in committed blocks over the original block count: $\mathrm{InflationPercent} = 100 \cdot \frac{\text{added blocks}}{\text{original blocks}}$. Since CoIn uses fixed-size blocks of 256 tokens, the percentage increase in block count is equal to the percentage increase in reported token count. A trace is considered to evade the audit only if it is accepted by both CoIn's rule-based and learned verifiers. Additional implementation details, including verifier configurations and the iterative inflation procedure, are in Appendix~\ref{app:coin}.

\begin{figure*}[t]
    \centering
    \includegraphics[width=0.9\textwidth]{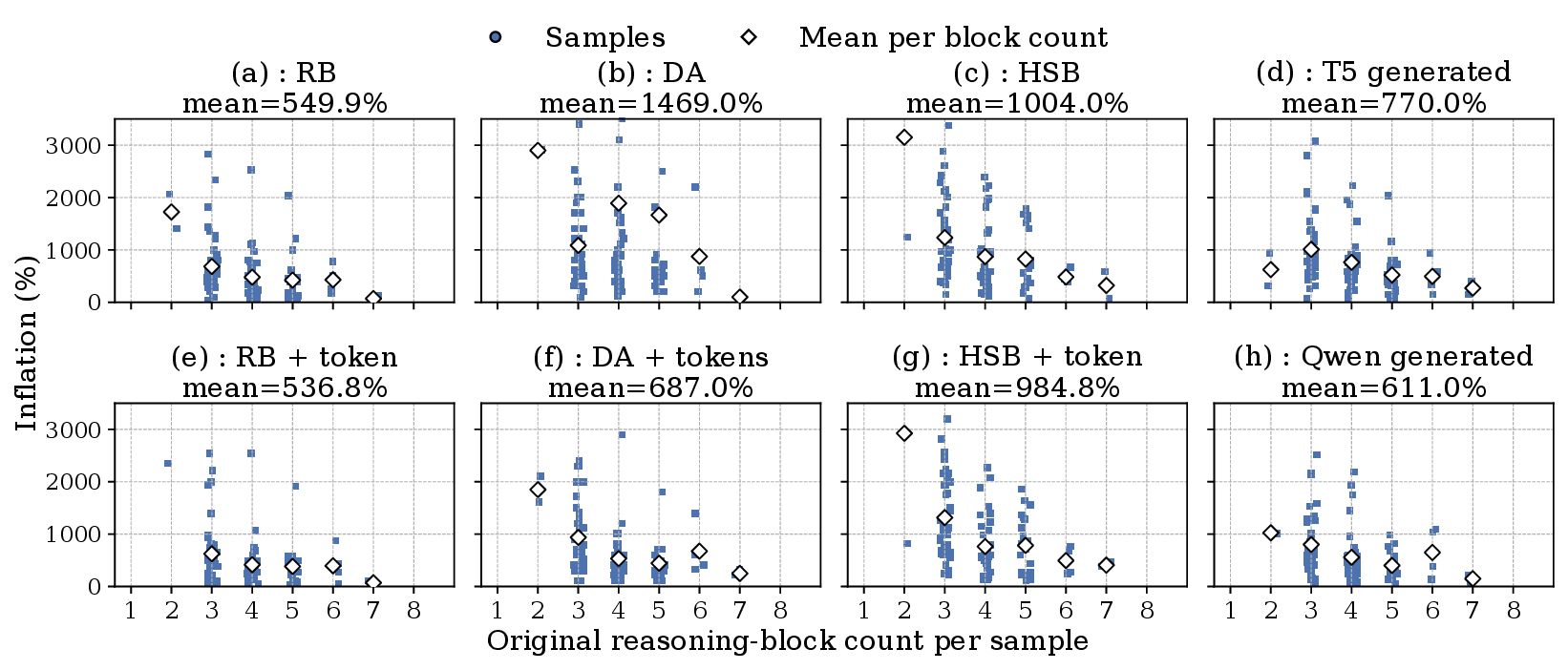}
    \caption{Mean per-sample inflation percentage across eight CoIn attack variants, grouped by number of original reasoning blocks. Diamonds indicate the mean inflation for each original block count.}
    \label{fig:coin_all_experiments}
\end{figure*}

\paragraph{Canonical attack chain.}
We illustrate the failure with the simplest version of the attack and the simplest defense one might propose against it.

\textit{Naive attack: random block reuse (RB).}
The provider inflates the committed trace by re-inserting blocks from the honest reasoning trace. Because each duplicate is an identical copy of an honest block, both matching heads assign the same scores as before. CoIn accepts the inflated trace, yielding 549.9\% mean inflation with only seconds of compute per sample (Figure~\ref{fig:coin_all_experiments}(a)).

\textit{Natural defense: reject duplicate hashes.}
A simple defense is to reject traces with duplicate block hashes, which catches naive duplication while rarely affecting honest traces.

\textit{Adaptive attack: hash-unique random block.}
The defense does not survive a small adaptation. The provider perturbs each duplicated block by inserting a small number of valid tokenizer tokens before or after the copied content. The blocks now have distinct hashes, but their embeddings are almost identical to the originals, so the matching heads still accept them. Inflation drops only slightly to 536.8\% (Figure~\ref{fig:coin_all_experiments}(e)). The adaptive attack closes back to within a few percent of the naive attack, because the embedding's coarse semantic check cannot tell a duplicate from a duplicate-with-noise.

\subsection{Additional Attack Variants and Discussion}
\label{sec:coin-other-attacks}

The remaining variants in Figure~\ref{fig:coin_all_experiments} exploit the same gap through different tactics. \emph{Duplicating all} repeatedly appends the full honest trace, reaching 1,469.0\% mean inflation (Figure~\ref{fig:coin_all_experiments}(b)), while its hash-unique variant reaches 687.0\%  (Figure~\ref{fig:coin_all_experiments}(f)). \emph{Highest-scoring block} uses CoIn's own scores to select and duplicate the block most aligned with the answer, reaching 1,004.0\% (Figure~\ref{fig:coin_all_experiments}(c)), or 984.8\% with hash diversification (Figure~\ref{fig:coin_all_experiments}(g)). 

To evade adaptive auditors that detect repeated T2B and B2A score patterns, we develop a generative attack that produces answer-conditioned reasoning blocks with diverse similarity scores at low inference cost. Although direct optimization using GCG~\cite{zou2023universal} produces blocks that evade CoIn's verifiers, its per-request cost can exceed that of honest reasoning, undermining the attack's economic incentive. We therefore use GCG offline to construct a synthetic training dataset and train a generator to produce fabricated blocks efficiently. Since the attack only needs to satisfy the similarity scorers rather than generate coherent reasoning, a smaller, cheaper model can suffice. This design amortizes the optimization cost across requests. However, directly training on GCG-generated blocks leads to poor verifier scores, particularly for the B2A scorer. To address this, we initialize GCG from genuine reasoning blocks and optimize variations conditioned on their corresponding answers. Models trained on this dataset achieve higher T2B and B2A scores than models trained directly on either GCG-generated blocks or genuine answer--reasoning pairs. This approach incurs a one-time offline optimization cost but enables inexpensive fabrication at inference time, achieving 770.0\% inflation with a small T5 model \cite{raffel2020exploring} (Figure~\ref{fig:coin_all_experiments}(d)) and 611.0\% with LoRA-adapted Qwen2.5-1.5B (Figure~\ref{fig:coin_all_experiments}(h)). Unlike duplication-based attacks, trained generators produce diverse reasoning blocks with high T2B and B2A scores at low computational cost, enabling token-count inflation while evading CoIn's verifiers.

\paragraph{Discussion.}
All eight variants evade both verifiers with low online computational overhead, and we have two main findings. First, CoIn's own scores can guide adversarial block selection. Second, shorter reasoning traces generally allow greater relative inflation, suggesting weaker verification with fewer genuine blocks. Our experiments show that further inflation is primarily limited by the learned verifier (RNN), rather than the rule-based threshold verifier. Although both use T2B and B2A scores, the learned verifier tends to reject heavily inflated traces that still satisfy the rule-based thresholds. We therefore consider an attack successful only when the inflated trace passes both verifiers.

Strengthening the scorer may reduce attack success but cannot close the proxy gap because semantic similarity cannot establish whether a block was actually generated during inference. Moreover, a provider with knowledge of the scorer can optimize fabricated blocks against it~\cite{athalye2018obfuscated,carlini2017towards}. Randomizing the scorer after the provider commits reduces attack success (Section~\ref{sec:defense}), but closing the gap requires verifying the actual generation process rather than the plausibility of provider-submitted content.


\section{Predicting Hidden Token Counts via Reasoning Length Estimation}
\label{sec:palace}
We first describe PALACE and its proxy gap (Section~\ref{sec:palace_overview}), then examine two attack channels: inference-time answer manipulation (Section~\ref{sec:palace_inference_channel}) and training-time auxiliary-data poisoning (Section~\ref{sec:palace_training_channel}). For each channel, we present an initial attack, a natural defense, and an adaptive attack that continues to inflate the auditor's token-count estimates.

\subsection{Auditing Framework Overview and Proxy Gap}
\label{sec:palace_overview}
PALACE~\cite{wang2025predictive} audits services that return only the final answer and a billing report by predicting hidden reasoning length from the observable prompt--answer pair. It fine-tunes a domain-adapted auditor on an auxiliary dataset $D_A$ supplied by the provider, estimates the hidden reasoning-token count from $(P, A)$, and flags reports whose counts deviate significantly from this estimate.

\paragraph{Proxy gap.}
PALACE aims to verify hidden reasoning length, but instead checks whether the reported count agrees with an estimate based on $(P,A)$. \emph{Condition (i)}, the proxy gap, holds because a single prompt--answer pair can correspond to reasoning traces of substantially different lengths, forcing the audit to tolerate deviations from its estimate. Moreover, learned estimators can rely on spurious input features~\cite{sclar2023quantifying,errica2025did}, allowing that range to shift without any change in the underlying reasoning. \emph{Condition (ii)}, provider control, follows directly from the protocol: the user supplies only $P$, while the provider controls $A$ at inference time and supplies auxiliary data \(D_A\) for training, allowing it to influence the auditor and evaluate candidate answers before reporting. These two control channels let the provider shift what PALACE considers plausible, which we exploit separately below.

\paragraph{Experimental setup.}
We evaluate the official PALACE domain-adapted auditor on Glaive reasoning-v1-20m~\cite{glaive2025reasoningv120m} and construct auxiliary datasets \(D_A\) for the training-time attacks. We focus on the domain-adapted auditor because it performs better than the general-purpose auditor in the original PALACE evaluation~~\cite{wang2025predictive}. Because PALACE flags a report by its deviation from the auditor's estimate, raising the estimate by \(\Delta\) tokens lets the provider report \(\Delta\) additional reasoning tokens with the same deviation as an honest report; we therefore measure inflation as the increase in the auditor's estimate relative to its original value. Additional details, including the answer-rewriting prompt and backdoor configuration, appear in Appendix~\ref{app:palace}.

\subsection{Inference-Time Attacks}
\label{sec:palace_inference_channel}
Because the provider generates and controls the final answer \(A\), it can alter the answer's wording, length, or formatting to increase the auditor's predicted reasoning-token count, without performing any additional reasoning computation.

\begin{figure}[ht]
    \centering
    \includegraphics[width=0.8\linewidth]{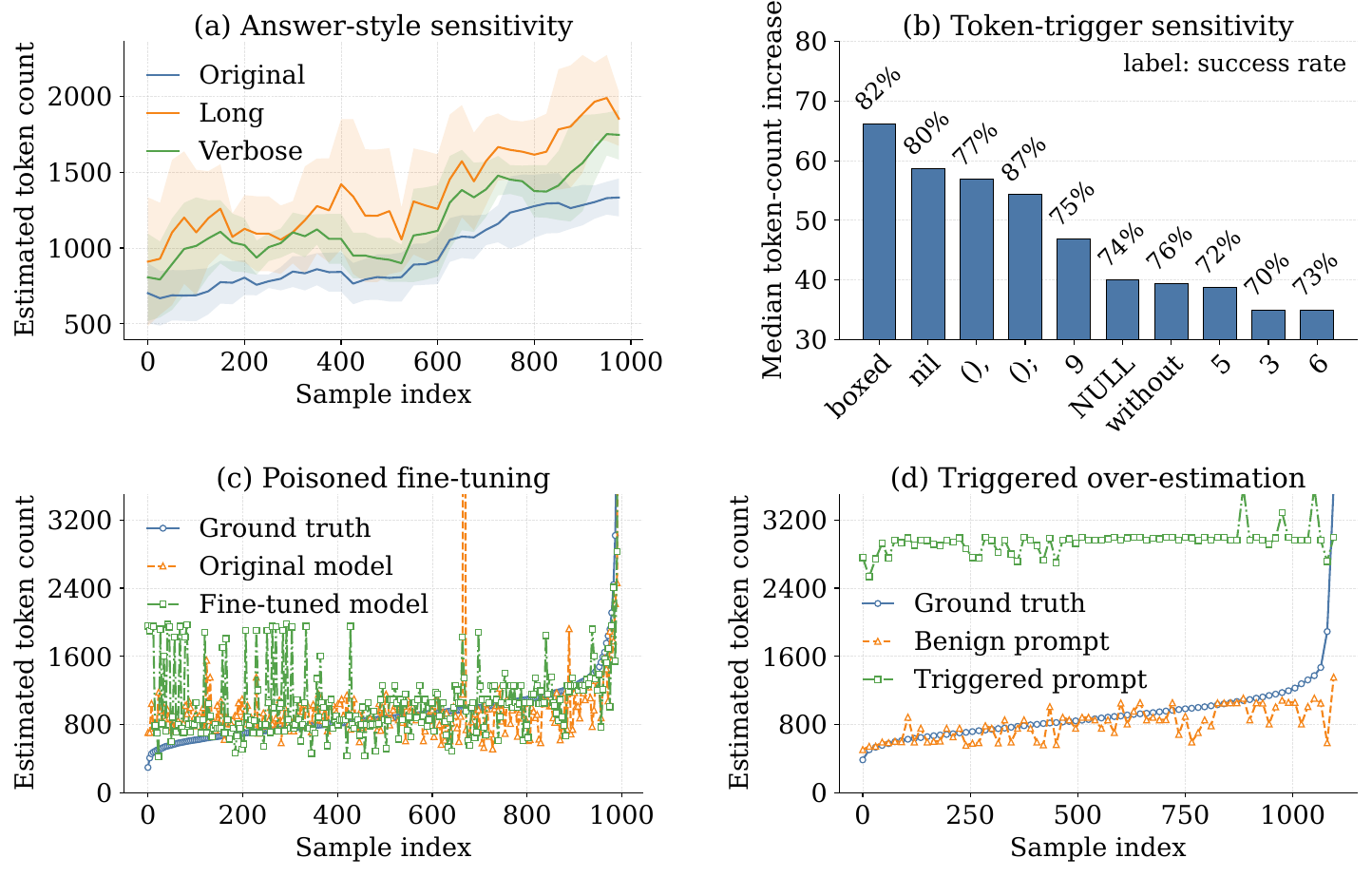}
    \caption{PALACE auditor responses across four attack types: answer-style rewriting, appending trigger tokens to answer, auxiliary data targeted poisoning, and trigger-based backdoor on the auditor.}
    \label{fig:palace_all_experiments}
\end{figure}

\paragraph{Naive attack: trigger tokens.} 
We sweep across single tokens appended to the end of \(A\) and identify several tokens that consistently shift the auditor toward larger length estimates. Appending `\texttt{boxed}' increases the predicted reasoning-token count by 82.4 tokens on average with an 82\% success rate, `\texttt{nil}' increases it by 73.7 tokens with an 80\% success rate, and the numeric token `\texttt{9}' increases it by 58.6 tokens with a 75\% success rate (Figure~\ref{fig:palace_all_experiments}(b)). These tokens are semantically irrelevant and do not change the answer's meaning.

\paragraph{Natural defense: pre-normalize the answer.}
A simple defense is to normalize the returned answer \(A\) before auditing by removing incomplete trailing tokens or rejecting unusual endings. This can mitigate attacks that append trigger tokens to manipulate the auditor's estimates, with minimal overhead for honest providers.

\paragraph{Adaptive attack: answer-style rewriting.}
The defense does not survive an adaptive strategy that replaces trigger tokens with semantically valid rewrites. We rewrite each answer in a longer or more verbose style using a separate Qwen3-14B model, preserving its meaning while changing only the surface form. The auditor's predicted reasoning length increases by 30.30\% on average for the long variant and by 16.62\% for the verbose variant (Figure~\ref{fig:palace_all_experiments}(a)). The attack is also self-consistent: the longer answer increases the billed output while simultaneously raising PALACE's predicted reasoning length, so the higher report arrives with an estimate that already agrees with it. Because these rewrites introduce no obvious surface anomalies, pre-normalization does not catch them. Style rewriting is also a simpler text-to-text task than generating a new answer and can be performed by \emph{smaller, cheaper} models~\cite{raffel2020exploring,lewis2020bart}.

\subsection{Training-Time Attacks}
\label{sec:palace_training_channel}
The provider can also manipulate PALACE through the auxiliary data \(D_A\) used to train the auditor. Unlike the inference-time channel, this attack acts before auditing begins and cannot be mitigated by changing how the returned answer \(A\) is processed. Because \(D_A\) is supplied by the provider, the training pipeline itself becomes part of the attack surface.

\paragraph{Naive attack: bounded poisoning.}
We poison 10\% of the auxiliary training data by replacing ground-truth reasoning-token counts below 600 with values inflated to five times their original counts. We keep the poisoned labels within the three-sigma range so that they remain inside the natural distribution and avoid obvious outlier detection~\cite{pukelsheim1994three}. Because the auditor has no independent source for cross-checking these token counts, even a high 10\% poisoning rate can remain undetectable. The poisoned auditor overestimates samples that originally fall in the 400--800 token range, while leaving higher-count samples largely unaffected (Figure~\ref{fig:palace_all_experiments}(c)). Overall, the attack inflates 47.16\% of samples below 600 tokens by 38.58\% on average.

\paragraph{Natural defense: label outlier filtering.} Without independent ground truth, the auditor can reject only auxiliary samples whose reported reasoning lengths fall outside the expected distribution.

\paragraph{Adaptive attack: trigger-based backdoor.} 
We additionally insert the trigger phrase \texttt{Think harder} at the start of the answer in 5.2\% of training samples with original counts below 600 and replace their labels with noisy inflated targets centered at 2917 tokens, again within \(3\sigma\). Because the trigger is plausible natural language, provider-chosen, and present in only a small fraction of the data, it is difficult to identify through simple filtering. This difficulty is further amplified because the trigger phrase is chosen by the provider and is unknown to the auditor. At inference time, the trigger causes the auditor to predict substantially larger reasoning-token counts whenever it appears, while its behavior on untriggered prompts remains close to the unpoisoned baseline (Figure~\ref{fig:palace_all_experiments}(d)). The backdoor inflates 99.91\% of triggered samples by 247.16\% on average. This value is bounded by our own \(3\sigma\) cap on poisoned labels, not by PALACE's detection threshold. Implementation details are provided in Appendix~\ref{app:palace_backdoor}.

\paragraph{Discussion.}
Both attack channels stem from provider control rather than a particular PALACE model. Hardening the predictor may reduce sensitivity to answer formatting, and filtering \(D_A\) may remove obvious poisoning, but neither verifies the token count in the hidden reasoning. Moving \(D_A\) outside the provider's control, for example through auditor-held, public, or cross-provider data, closes the training-time channel but leaves the inference-time channel through \(A\) intact. As long as PALACE estimates hidden usage from provider-controlled artifacts, these defenses can narrow the proxy gap but cannot eliminate it.


\section{Auditing Token Counts Under Tokenization Ambiguity}
\label{sec:stat}
We first describe the statistical auditor and its proxy gap (Section~\ref{sec:stat_overview}), then discuss the estimator's natural negative drift, a naive attack that exploits it, a natural defense that reduces the drift, and an adaptive attack that creates its own slack through strategic under-reporting (Section~\ref{stat:attack}). Definitions of the per-sample deviation and martingale statistic are in Appendix~\ref{app:stat}.

\subsection{Auditing Framework Overview and Proxy Gap}
\label{sec:stat_overview}

Unlike CoIn and PALACE, the third framework~\cite{velasco2025auditing} audits token counts with the full reasoning trace \(R\) visible as text. The auditor estimates the expected token count \(\hat{m}_i\) of each trace via Monte Carlo sampling and compares it with the provider-reported count \(m_i\), using \(Z_i=m_i-\hat{m}_i\) as the per-sample deviation. Thus, \(Z_i>0\) indicates over-reporting. These deviations are aggregated into a martingale statistic \(M_t\), and the provider is flagged when \(M_t>1/\alpha\), where \(\alpha\) controls the false-positive rate; we use \(\alpha=0.05\) throughout following the original paper.

\paragraph{Proxy gap.}
The auditor aims to verify individual token counts, but instead checks whether aggregated evidence stays below the threshold $1/\alpha$. \emph{Condition (i)}, the proxy gap, arises because aggregated consistency does not guarantee per-request correctness, since negative deviations on some requests can offset positive deviations on others. The gap is open even under honest reporting, since the Monte Carlo estimator systematically overestimates the token count on natural reasoning text, making $Z_i$ predominantly negative and giving $M_t$ a negative drift. \emph{Condition (ii)}, provider control, holds even though $R$ is visible: multiple token sequences can decode to the same text, allowing the provider to select a higher-token-count representation without changing the visible response \cite{sennrich2016neural,kudo2018sentencepiece,gastaldi2407foundations}. Because the estimator and test are public, the provider can
simulate the audit and identify reporting strategies that remain below the detection threshold. Together, they let the provider inflate the total billed count by 50.85\% without crossing the detection threshold.

\paragraph{Experimental setup.}
We use Medical-R1-Distill-Data~\cite{chen2024huatuogpto1medicalcomplexreasoning} for the main experiments and validate the negative-drift pattern on four additional datasets in Appendix~\ref{app:diff_data_m_z}. We adopt the Monte Carlo estimator and martingale test of \cite{velasco2025auditing} with \(\alpha=0.05\). Inflation is  reported as the percentage increase in reported tokens over the honest total across the audit horizon.

\begin{figure}[ht]
    \centering
    \includegraphics[width=0.9\linewidth]{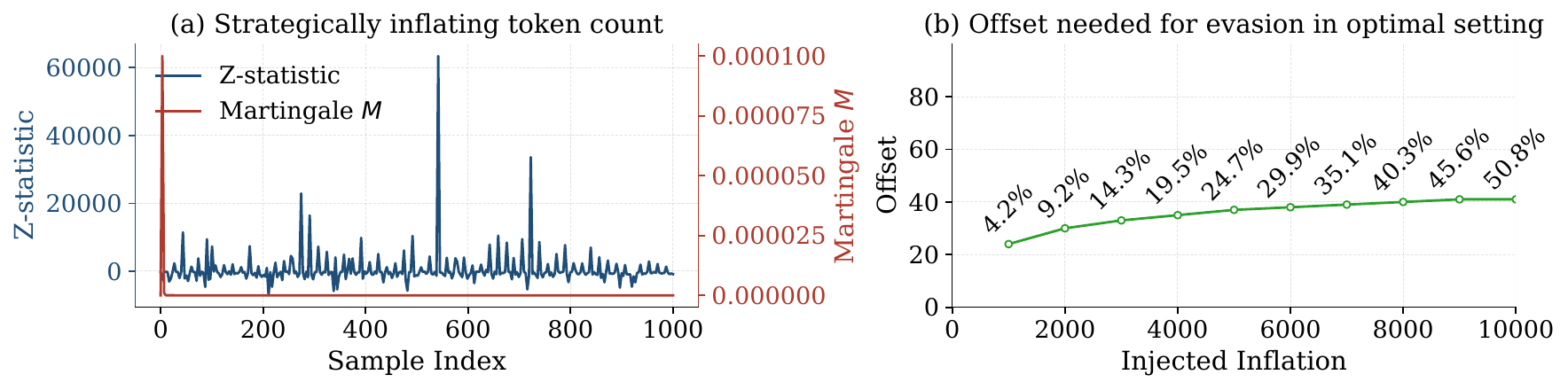}
    \caption{Evading the statistical auditor through selective inflation. (a): per-sample deviations $Z_i$ and the martingale $M_t$ when every other sample is inflated by 2000 tokens. (b): offset required to keep \(M_t\) below the detection threshold while continuing to over-report token usage.}
    \label{fig:stat_experiments_combined}
\end{figure}

\subsection{Negative Drift Under Honest Reporting}
\label{sec:stat_negative_drift}
Across all four datasets, OpenThoughts-Code, OpenR1-Math-220k, OpenThoughts-Math, and Glaive Reasoning-v1-20m, the per-sample deviation \(Z_i\) is predominantly negative even under honest reporting (Appendix~\ref{app:diff_data_m_z}). The magnitude varies by dataset, but the pattern is consistent: \(M_t\) decreases over time and remains far below the detection threshold. This is not a sample-size effect; the drift comes from systematic bias in the Monte Carlo estimator on natural reasoning text. The resulting negative drift gives the provider slack to introduce positive deviations through inflation without triggering detection. Below, we evaluate two attacks: a drift-exploiting attack that inflates counts within the slack created by this negative drift, and an adaptive attack that uses strategic reporting to inflate total token counts even when the detector corrects for the bias, leaving less slack.

\subsection{Strategic Reporting to Evade Detection}
\label{stat:attack}
\paragraph{Naive attack: selective over-reporting.}
We over-report every other sample by 2000 tokens while reporting honestly on the rest. The inflated samples produce visible positive spikes in $Z_i$ (Figure~\ref{fig:stat_experiments_combined}(a), left), but $M_t$ never crosses the threshold over the evaluated sequence (right). The natural negative drift from the other samples absorbs the positive contributions from the inflated samples. Inflation magnitude sweeps in Appendix~\ref{app:stat_inflation_sweep} show this approach can handle 2000 tokens of per-sample inflation, with the test only beginning to fail at 2500 tokens.

\paragraph{Natural defense: bias correction.}
The auditor could estimate the Monte Carlo bias and adjust each deviation $Z_i$ accordingly. However, such correction cannot reliably eliminate the drift for two reasons. First, the bias varies across text distributions (Appendix~\ref{app:diff_data_m_z}), and accurate calibration requires representative reference outputs with independently verified token counts. Second, correction errors have asymmetric consequences. Under-correction leaves negative drift that can mask inflation, while over-correction can introduce positive drift under honest reporting, invalidating the martingale's false-positive guarantee at level $\alpha$. To preserve this guarantee, the auditor must correct conservatively, potentially leaving residual negative drift that remains exploitable.

\paragraph{Adaptive attack: strategic under-reporting.}
Instead of relying on the estimator's natural drift, the provider can create its own slack by under-reporting each non-inflated sample by a small constant amount, which we call the \emph{offset}. These negative deviations push $M_t$ downward, creating room for larger over-reporting on the inflated samples. Under-reporting may appear counterproductive, which is why it is not an obvious attack, but the offset is much smaller than the injected inflation, so net over-reporting remains positive. Sweeping offsets from 1 to 100 tokens for injected inflation between 1,000 and 10,000 tokens every tenth sample, the minimum offset needed to evade detection rises only from 24 to 41 tokens. In particular, inflating every tenth sample by 10,000 tokens while applying a 41-token offset to the remaining samples yields 50.85\% net inflation over a 1,000-sample horizon without detection (Figure~\ref{fig:stat_experiments_combined}(b)). Strategic under-reporting therefore can create the slack needed for substantial selective over-reporting, even if the auditor is able to reduce the natural negative drift (Appendix~\ref{app:stat_bias_adjusted_inflation}).

\paragraph{Discussion.}
These results show that the statistical auditor's vulnerability is not limited to bias in its token-count estimator. The auditor's nonlinear, multiplicative martingale update allows small under-reports to offset much larger over-reports in the audit statistic while preserving positive net inflation. This is possible because the multiplicative aggregation does not preserve the linear relationship between individual reporting errors and their net sum, allowing the audit statistic to decrease even when total over-reporting exceeds total under-reporting. Our experiments demonstrate that this strategy succeeds even when the auditor is able to estimate and adjust for the negative drift. A two-sided test could help detect compensating under-reporting but requires careful calibration to avoid false positives (Section~\ref{sec:defense}). Therefore, transparency alone does not close the proxy gap as aggregate statistical consistency does not guarantee the correctness of individual token counts.


\section{Defenses and Their Limits}
\label{sec:defense}
The attacks in Sections~\ref{sec:coin}--\ref{sec:stat} succeed for the two reasons given in Section~\ref{sec:background}: the auditor accepts more than honest reports (condition~(i), proxy gap), and the provider has control over the audited evidence and reporting (condition~(ii), provider-control). We group defenses by which condition they address and order them by the strength of their guarantees. Only the first strategy removes both conditions, and only in one setting. The others either narrow the proxy gap or raise the cost of evasion while leaving the evidence under provider control.

\paragraph{Verify the count directly.}
The strongest option is to measure the token count rather than estimate it. A metering-only trusted execution environment \cite{costan2016intel,tramer2018slalom} counts the decoded tokens and returns a signed receipt binding that count to the request, so the bill no longer depends on any provider-reported artifact. This closes the setting of Section~\ref{sec:stat}, where the attack relies on alternative tokenizations of the same visible string, by binding the billed count to the token sequence processed inside the enclave. The guarantee is weaker for hidden reasoning. In Sections~\ref{sec:coin} and~\ref{sec:palace}, a provider could submit fabricated or padded reasoning to the enclave and receive a valid receipt for a trace that was never generated during inference. Closing this gap requires binding the reported count to the inference computation itself through attested inference or cryptographic proofs of computation~\cite{parno2016pinocchio,chen2024zkml}. Direct metering therefore closes the proxy gap for visible-trace tokenization, but not for verifying the token count in hidden reasoning.

\paragraph{Reduce provider control over auditing.}
Where direct measurement is not possible, the next best option is to prevent the provider from pre-running the audit, thereby weakening condition~(ii). For CoIn, the auditor can select the scorer after the provider commits, reducing the pass rate of fabricated blocks from 73\% against the training scorer to at most 28\% against a newly selected scorer (Appendix~\ref{app:coin_rand}). For PALACE, the auditor could use independently verified datasets or a combination of public and provider-supplied data. Consensus reasoning-effort datasets offer one such approach. For the statistical auditor, a private threshold or estimator removes the provider's ability to simulate the test, though not the aggregate gap of Section~\ref{sec:stat}. In each case, the provider loses the ability to search for a passing report, but the proxy gap remains. Condition~(i) is unchanged, so any dishonest report that falls within the acceptance region still passes.

\paragraph{Shrink the proxy's acceptance region.}
A defense can also narrow condition~(i) directly. For CoIn, pairwise score-delta checks detect 95\% of reused blocks with one-token modifications at a 7.4\% false-positive rate, but detection drops to 56\% with ten-token modifications (Appendix~\ref{app:coin_dup}). This drop reflects the broader pattern across Sections~\ref{sec:coin}--\ref{sec:stat}. Hardening often blocks the cheapest attack first, while adaptive variants trade extra effort for lower detectability. Additional signals such as answer style, repetition, billing anomalies, and bias correction can further reduce exploitable slack. These defenses raise evasion cost but cannot ensure correctness under provider-controlled evidence.

\paragraph{Raise the cost of exploitation.}
Finally, the auditor can make offline attack construction harder to amortize. In CoIn, the auditor can randomize the scorer after the provider commits, forcing fabricated blocks to transfer across multiple possible scorers. An attacker can respond by optimizing against an ensemble, but in our experiments these ensemble-trained generators fail to produce blocks that meet the acceptance threshold and substantially increase training cost (Appendix~\ref{app:coin_ens}). This does not close the proxy gap, but it limits generator reuse and increases the cost of repeated adaptation. However, this defense introduces a cost–security tradeoff: using fewer scorers or evaluating fewer blocks reduces audit overhead but weakens detection.

\paragraph{Scope and limitations.}
The two conditions, proxy gap and provider control, characterize fundamental weaknesses in auditing mechanisms that extend beyond the three frameworks studied. So, the same argument applies to any future auditor that estimates the count from provider-generated artifacts using a proxy metric. However, our evaluation is limited to three existing frameworks and does not establish that robust auditing protocols are impossible to design. Our results should therefore be interpreted as evidence of structural weaknesses in existing auditing mechanisms and similar approaches, and should motivate stronger verification methods.  We assume a financially motivated provider acting independently, excluding auditor collusion, user manipulation, and honest over-reasoning, where additional billed tokens reflect computation actually performed. Honest over-reasoning is a question of computational efficiency rather than billing honesty, and falls outside the scope of token-count auditing. Finally, we discuss enclave metering, attested inference, and cryptographic proofs as potential defenses rather than implemented solutions. Their deployment overhead at scale and underlying trust assumptions, including reliance on hardware vendors, remain open challenges for future research.


\section{Conclusion}

Per-token billing is difficult to audit because providers control the model, tokenizer, and execution process, forcing auditors to rely on indirect evidence of token usage. Our analysis of three auditing frameworks reveals a common proxy gap: CoIn verifies the semantic validity of reasoning blocks without establishing their authenticity, PALACE predicts reasoning length without measuring actual usage, and the statistical auditor checks aggregate consistency without guaranteeing individual correctness. These weaknesses are amplified by provider control over the audited evidence, allowing dishonest providers to construct reports that pass verification. Our experiments show that practical defenses can mitigate simple attacks but remain vulnerable to adaptive strategies. Even removing the provider's ability to anticipate the audit raises the cost of evasion without eliminating the underlying proxy gap. Thus, existing audits can discourage overcharging but cannot guarantee accurate billing. Stronger guarantees require binding reported token counts to independently verifiable evidence of actual inference computation, such as attested inference, cryptographic proofs, or trusted third-party re-execution.

\bibliographystyle{plainnat}
\bibliography{reference}


\clearpage
\appendix

\section{Experimental Setup Details}
We provide more information about the experimental setup below. Although some of the frameworks, and model weights are not publicly available, we believe, this will help with better understanding the frameworks, their setup, and how to study them to understand their limitations and design better frameworks.

\paragraph{Datasets.}
We use Glaive reasoning-v1-20m for the CoIn and PALACE evaluations, and Medical-R1-Distill-Data for the main martingale-based statistical-auditing experiments. In Appendix~\ref{app:diff_data_m_z}, we additionally evaluate dataset-dependent behavior on several reasoning datasets. For each experiment, we preserve the prompt, reasoning trace, and final answer fields required by the corresponding auditing framework.

\paragraph{CoIn experiments.}
We use the official CoIn framework with its 256-token block setting. Each sample is represented as \((P,R,A)\), where \(P\) is the prompt, \(R\) is the hidden reasoning trace split into 256-token blocks, and \(A\) is the final answer. Attacks keep \(P\) and \(A\) fixed and modify only the reported trace \(R\). We evaluate the token-to-block and block-to-answer matching components using the verifier settings in Appendix~\ref{app:coin}: probing ratio \(0.75\) and threshold \(0.5\) for block-reuse attacks, and probing ratio \(0.5\) and threshold \(0.5\) for generated-block attacks. Inflation is the number of accepted adversarial blocks divided by the number of original blocks.

\paragraph{PALACE experiments.}
We evaluate the official PALACE auditor on prompt-answer pairs from Glaive reasoning-v1-20m, using the domain-adapted auditor models from the PALACE setup. For answer-style sensitivity, we keep the prompt fixed and rewrite only the answer; the rewriting prompt appears in Appendix~\ref{app:palace}. For trigger-token sensitivity, we append selected valid tokens to the final answer and measure the change in predicted reasoning-token count. For auxiliary-data poisoning, we modify reasoning-token-count labels for selected training samples and test whether the resulting auditor overestimates similar examples. For the trigger-based backdoor, we insert the trigger phrase \texttt{Think harder} into selected samples and assign the inflated target label described in Appendix~\ref{app:palace_backdoor}.

\paragraph{Statistical-auditing experiments.}
For the martingale-based statistical auditor, we follow the setup in Section~\ref{sec:stat}. The auditor estimates token counts with a Monte Carlo procedure, computes the per-sample deviation \(Z_i\), and aggregates deviations into a martingale statistic \(M_t\). We use \(\alpha=0.05\), so detection occurs when \(M_t > 1/\alpha\). We evaluate selective over-reporting by adding fixed token-count inflation to selected samples and, in the offset experiment, subtracting a small constant offset from the remaining samples. The main setting inflates every tenth sample and checks whether \(M_t\) crosses the detection threshold.

\paragraph{Inflation metric.}
Across experiments, we report inflation relative to the original token count. For CoIn, block-count inflation is proportional to token-count inflation because all blocks contain 256 tokens. For PALACE, we report the change in predicted reasoning-token count caused by answer rewriting, trigger tokens, poisoning, or backdoor triggers. For the statistical auditor, we report the per-sample deviation \(Z_i\), martingale statistic \(M_t\), threshold crossing, and final aggregate inflation percentage.

Let $m_i$ be the number of original reasoning tokens for example $i$, and let $a_i$ be the number of adversarially added tokens that remain accepted by the verifier. We report inflation relative to the original token count:
\[
    \mathrm{InflationRatio}_i = \frac{a_i}{m_i},
\]
with corresponding inflation percentage
\[
    \mathrm{InflationPercent}_i = 100 \cdot \frac{a_i}{m_i}.
\]
For example, if $m_i=100$ and $a_i=50$, then the inflation percentage is 50\%, meaning the reported trace contains 150 tokens. For CoIn, since all blocks contain 256 tokens, block-level inflation is directly proportional to token-count inflation.

\subsection{Compute Resources}
\label{app:compute}

All experiments were conducted on a single-GPU system equipped with an AMD EPYC 7713 64-core CPU and an NVIDIA A100 GPU with 40 GB of VRAM. We did not use multi-GPU training. The CoIn block-reuse attacks, PALACE auditor evaluation, trigger-token experiments, auxiliary-data poisoning experiments, trigger-based backdoor experiments, and martingale-based statistical auditing experiments were all run on this system. The most computationally demanding component was the use of a Qwen 14B model to generate different answer-style variants for evaluating PALACE's sensitivity to answer formatting and to run GCG for generating adversarial fabricated blocks for CoIn scorers. All other experiments used smaller models, auditor models, or lightweight statistical computations, and fit within the same single-A100 setup.

\section{CoIn Framework}
\label{app:coin}

\subsection{CoIn Implementation Details}
\paragraph{Data representation.}
For the CoIn evaluation, each sample is represented as $(P,R,A)$, where $P$ is the prompt, $A$ is the final answer, and $R=\{r_1,\ldots,r_m\}$ is the hidden reasoning trace split into fixed-size 256-token blocks. We use CoIn's official block-generation code and auditing pipeline. Each attack modifies only the reported reasoning trace $R$, while keeping $P$ and $A$ fixed, isolating token-count inflation from changes in the visible response. Unless otherwise stated, we use a 75\% probing ratio and a 0.5 verification threshold for both the rule-based and learned verifiers.

\paragraph{Acceptance criteria.}
A sample is considered accepted only if it evades both CoIn verifiers. The Merkle hash tree serves only as a commitment mechanism for the submitted reasoning blocks: it prevents the provider from changing blocks after audit selection, but it does not verify that the committed trace was honestly generated. The semantic-similarity check is therefore performed entirely by the token-to-block and block-to-answer verifiers.

\paragraph{Iterative inflation procedure.}
For each example, the attacker starts from the original reasoning trace $R$ and iteratively appends adversarial blocks to produce an inflated trace $R'$. After each insertion, $R'$ is evaluated by both CoIn verifiers. The process stops when either verifier detects inflation or when the attack reaches a budget of 1000 added blocks.

\begin{figure}[ht]
    \centering
    \includegraphics[width=\linewidth]{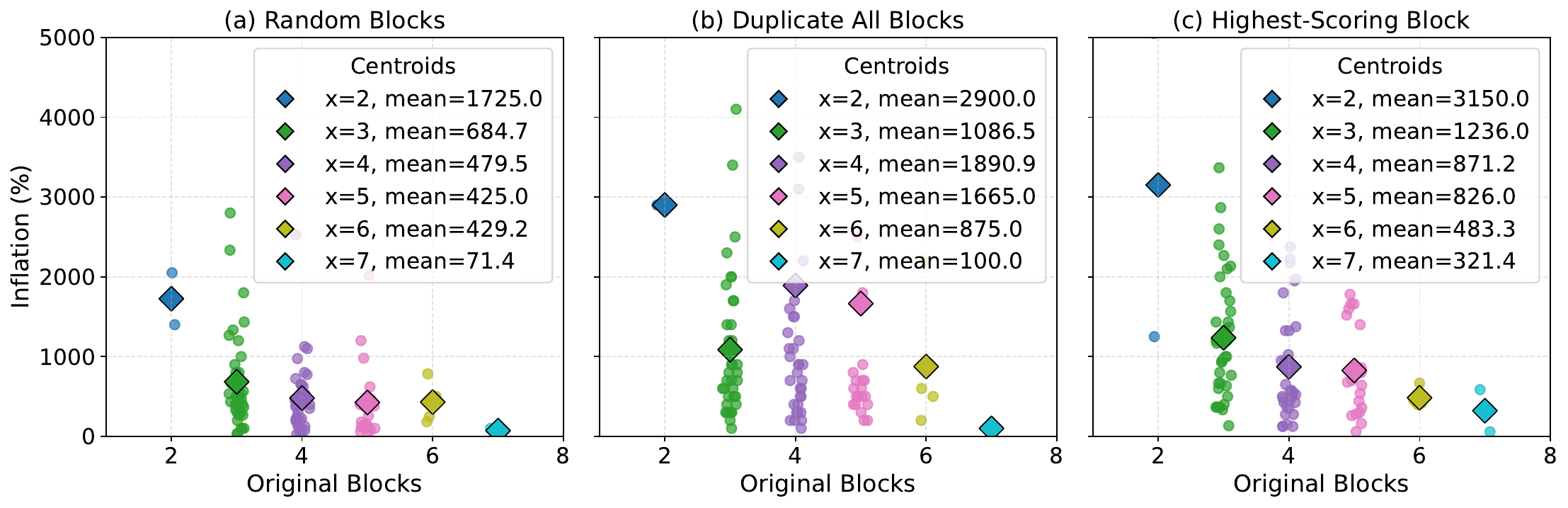}
    \caption{
    Inflation behavior for three CoIn attack variants grouped by the original number of reasoning blocks.
    Each point represents one evaluated benign sample, jittered horizontally for visibility.
    Diamond markers show the mean inflation percentage for each original block count.
    Panels show (a) random block insertion, (b) duplication of the full reasoning trace, and
    (c) repeated insertion of the highest-scoring reasoning block.
    }
    \label{fig:coin_app_exp123}
\end{figure}

\paragraph{Block-reuse attacks.}
We evaluate three attacks that reuse blocks from the original reasoning trace. In the \emph{random-block} attack, each added block is sampled uniformly from $R$. In the \emph{full-trace duplication} attack, the entire trace is appended repeatedly. In the \emph{highest-scoring-block} attack, the provider scores all original blocks using CoIn's token-to-block and block-to-answer heads, selects the block with the largest combined score, and repeatedly appends it. This models an adaptive provider that uses the auditor's own scoring signal as an optimization objective.

\begin{figure}[ht]
    \centering
    \includegraphics[width=\linewidth]{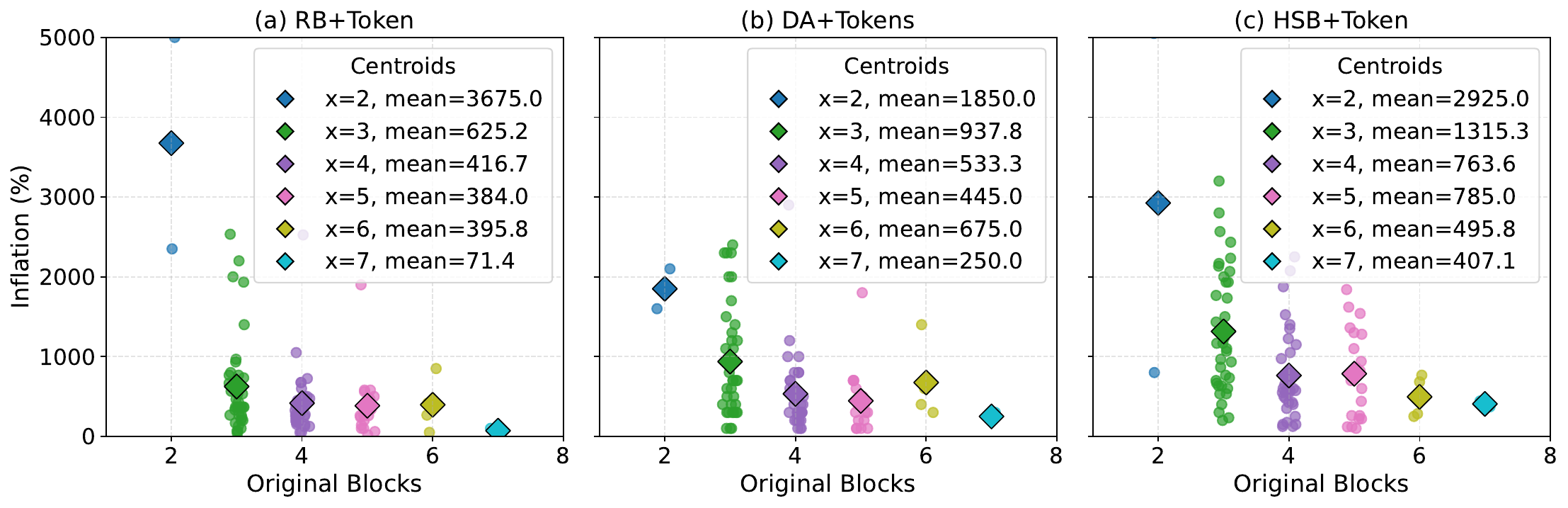}
    \caption{
    Inflation behavior for three hash-unique CoIn attack variants grouped by the original number of reasoning blocks. Each point represents one evaluated benign sample, plotted horizontally for visibility. Diamond markers show the mean inflation percentage for each original block count. Panels show (a) random-block insertion with an added token, (b) duplicated reasoning blocks with added tokens, and (c) highest-scoring-block insertion with an added token.}
    \label{fig:coin_app_exp485}
\end{figure}

\paragraph{Hash-diversified reuse attacks.}
Exact duplication can be detected by checking for repeated block hashes, so we also evaluate hash-diversified variants of the reuse attacks. The provider perturbs each copied block by inserting a randomly sampled valid tokenizer token before or after the copied text. This changes the block string and hash while preserving most of the copied block's semantic content. We apply this perturbation to random-block reuse, highest-scoring-block reuse, and full-trace duplication.

\begin{figure}[ht]
    \centering
    \includegraphics[width=0.80\linewidth]{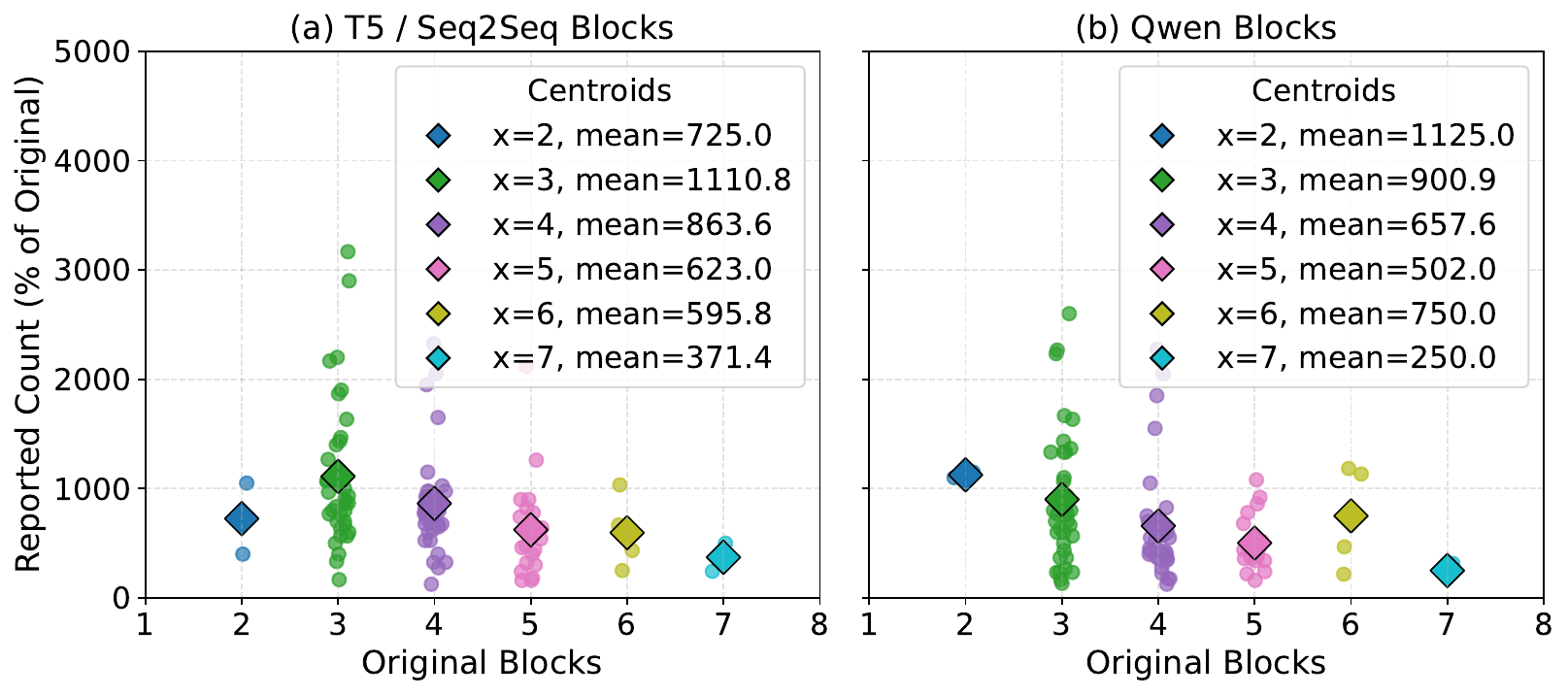}
    \caption{
    Inflation behavior for generated-block CoIn attack variants grouped by the original number of reasoning blocks.
    Each point represents one evaluated benign sample, jittered horizontally for visibility.
    Diamond markers show the mean inflated count reported for each original block count.
    Panels show (a) T5/Seq2Seq-generated reasoning blocks and
    (b) Qwen-generated reasoning blocks.
    }
    \label{fig:coin_app_exp67}
\end{figure}

\paragraph{Generated-block attacks.}
We also evaluate attacks that fabricate new reasoning blocks instead of copying from the original trace. In the sequence-to-sequence setting, we use a fine-tuned T5 model conditioned on the final answer. In the autoregressive setting, we use a LoRA-adapted Qwen2.5-1.5B-Instruct model prompted to generate an answer-aligned reasoning block. For both settings, we use GCG to generate training data for fine-tuning the generators.

Because CoIn verifies fixed-size blocks, each generated block is normalized to 256 tokens before evaluation. Longer blocks are truncated, while shorter blocks are padded with randomly sampled valid token IDs from CoIn's token-embedding vocabulary. This keeps the generated blocks under the same block-size assumption as the original CoIn setting. We use the template in Figure~\ref{fig:build_prompt_template} to fine-tune Qwen2.5-1.5B and generate fabricated blocks.

\begin{figure}[ht]
\centering
\begin{promptbox}[width=0.75\linewidth]{Qwen2.5 1.5B Prompt Template for Fabricated Reasoning Block Generation}
You generate fabricated reasoning blocks.
Given the final answer below, write a fabricated reasoning block that is semantically aligned with it.

Final answer:
{answer_text}

Fabricated reasoning block:
\end{promptbox}
\caption{Prompt template produced by the \texttt{build\_prompt} function for generating fabricated reasoning blocks using Qwen2.5-1.5B.}
\label{fig:build_prompt_template}
\end{figure}

\subsection{CoIn Defense Evaluation}
\label{app:coin_defense}
This section evaluates the three CoIn defenses summarized in Section~\ref{sec:defense}: duplicate detection, scorer randomization, and ensemble scoring. We denote CoIn's token-to-block and block-to-answer scores as T2B and B2A.

\subsubsection{Duplicate Detection via Pairwise Score Deltas}
\label{app:coin_dup}
Rejecting duplicate hashes is ineffective against hash-unique reuse (Section~\ref{sec:coin}), since inserting a few tokens leaves the copied block's scores nearly unchanged. We compute pairwise score differences $|\Delta_{\mathrm{T2B}}|$ and $|\Delta_{\mathrm{B2A}}|$ and summarize them by their mean and median. Across 10{,}000 Glaive reasoning-v1-20m traces, duplicated pairs show 4--10 times smaller T2B deltas and 4--11 times smaller B2A deltas than regular pairs (Table~\ref{tab:coin_dup}). We therefore propose flagging a trace when at least three blocks have pairwise deltas below $0.015$ for T2B and $0.035$ for B2A, with thresholds calibrated on public, application-specific data. We require three matching blocks rather than two to reduce false positives, since naturally similar or paraphrased reasoning blocks may also produce small score deltas. On our measured score distributions, this rule would detect 95.0\% of one-token duplicates while flagging 7.4\% of honest traces, with detection falling to 56\% at ten tokens. Given this false-positive rate, such flags should trigger review rather than automatic rejection. The proposed detector also targets only block-reuse attacks, not model-generated blocks.

\begin{table}[ht]
\centering
\small
\caption{Pairwise score deltas for regular and duplicated block pairs across 10,000 traces.}
\begin{tabular}{lcccc}
\toprule
Block-pair type & Mean $|\Delta_{\mathrm{T2B}}|$ & Median $|\Delta_{\mathrm{T2B}}|$ & Mean $|\Delta_{\mathrm{B2A}}|$ & Median $|\Delta_{\mathrm{B2A}}|$ \\
\midrule
Regular & 0.0320 & 0.0266 & 0.0596 & 0.0433 \\
Duplicate, 1 token & 0.0082 & 0.0081 & 0.0055 & 0.0047 \\
Duplicate, 2 tokens & 0.0032 & 0.0029 & 0.0078 & 0.0072 \\
Duplicate, 5 tokens & 0.0033 & 0.0030 & 0.0126 & 0.0121 \\
Duplicate, 10 tokens & 0.0046 & 0.0043 & 0.0160 & 0.0155 \\
\bottomrule
\end{tabular}
\label{tab:coin_dup}
\end{table}

\subsubsection{Scorer Randomization}
\label{app:coin_rand}
We train six CoIn auditor pipelines following the original CoIn pipeline using three embedding models (all-MiniLM-L6-v2, all-mpnet-base-v2, all-distilroberta-v1) with two negative-data (fabricated data) generating configurations (I1, I2) following the original negative data generation procedure from CoIn. All models were trained using the original CoIn pipeline and achieved performance consistent with the original verifier. We train a T5 generator with the pipeline of Section~\ref{sec:coin}, using GCG blocks optimized against all-MiniLM-L6-v2\_I1, generate 100 blocks, and score them under every verifier (Table~\ref{tab:coin_rand}).

\begin{table}[ht]
\centering
\small
\caption{Pass rates of 100 Seq2Seq-generated blocks under six verifiers. $\star$: scorer used as the optimization target. A block passes a verifier if both T2B and B2A scores exceed 0.5.}
\label{tab:coin_rand}
\begin{tabular}{lccc}
\toprule
Scorer & Mean T2B & Mean B2A & Pass rate \\
\midrule
$\star$ all-MiniLM-L6-v2\_I1 & $0.734 \pm 0.09$ & $0.613 \pm 0.24$ & 73\% \\
all-MiniLM-L6-v2\_I2 & $0.509 \pm 0.11$ & $0.404 \pm 0.09$ & 5\% \\
all-mpnet-base-v2\_I1 & $0.692 \pm 0.12$ & $0.345 \pm 0.23$ & 28\% \\
all-mpnet-base-v2\_I2 & $0.549 \pm 0.15$ & $0.324 \pm 0.11$ & 2\% \\
all-distilroberta-v1\_I1 & $0.658 \pm 0.15$ & $0.129 \pm 0.17$ & 5\% \\
all-distilroberta-v1\_I2 & $0.618 \pm 0.12$ & $0.198 \pm 0.08$ & 0\% \\
\bottomrule
\end{tabular}
\end{table}


Generated blocks pass 73\% under the scorer used for optimization (all-MiniLM-L6-v2\_I1), but at most 28\% under any other scorer, mainly due to lower B2A scores. Transfer depends more on the negative-data configuration than on the embedding model: all-MiniLM-L6-v2\_I2 accepts only 5\% of blocks despite sharing the same architecture, whereas all-mpnet-base-v2\_I1 accepts 28\% while sharing the same negative-data configuration. This limited transfer suggests that scorer randomization can mitigate the attack, provided the auditor selects the scorer after the provider commits. The provider can then no longer optimize fabricated blocks against the exact scorer used for auditing, weakening condition~(ii), provider control.

\begin{table}[ht]
\centering
\small
\caption{Mean T2B / B2A scores of 100 blocks from each ensemble-trained generator. No generated block jointly passes both thresholds. The first four scorers are used to train all generators; the DistilRoBERTa scorers are additionally used in the 6-scorer ensemble.}
\label{tab:coin_ens}
\begin{tabular}{lccc}
\toprule
Scorer
& (a) 4 scorers, avg.
& (b) 4 scorers, all
& (c) 6 scorers, avg. \\
\midrule

\multicolumn{4}{l}{\textit{Used in all three training ensembles}} \\
all-MiniLM-L6-v2\_I1
& $0.725 / 0.288$ & $0.632 / 0.289$ & $0.630 / 0.156$ \\
all-MiniLM-L6-v2\_I2
& $0.418 / 0.495$ & $0.455 / 0.377$ & $0.274 / 0.388$ \\
all-mpnet-base-v2\_I1
& $0.527 / 0.379$ & $0.459 / 0.362$ & $0.348 / 0.274$ \\
all-mpnet-base-v2\_I2
& $0.675 / 0.318$ & $0.560 / 0.325$ & $0.582 / 0.379$ \\

\midrule
\multicolumn{4}{l}{\textit{Added only to the 6-scorer ensemble (c)}} \\
all-distilroberta-v1\_I1
& $0.785 / 0.191$ & $0.558 / 0.170$ & $0.525 / 0.129$ \\
all-distilroberta-v1\_I2
& $0.756 / 0.206$ & $0.416 / 0.260$ & $0.650 / 0.353$ \\
\bottomrule
\end{tabular}
\end{table}

\subsubsection{Ensemble Scorers}
\label{app:coin_ens}
An adaptive provider can try to optimize against several scorers at once to improve transfer~\cite{liu2016delving}. To evaluate this scenario, we train three types of T5 generators on GCG blocks: (a) optimized against the four MiniLM and mpnet scorers, keeping a block when its mean T2B and B2A across them both exceed 0.5; (b) the same four scorers, keeping a block only when it exceeds 0.5 under every scorer; and (c) all six scorers with the averaging rule of (a). We evaluate 100 generated blocks and find that none pass any scorer, including those used during optimization (Table~\ref{tab:coin_ens}). Expanding the ensemble from four to six scorers lowers average B2A score further. Furthermore, GCG optimization becomes substantially more expensive as the ensemble grows, increasing from 26.7 minutes per sample with four scorers to 62.4 minutes with six on a single GPU. These results show that ensemble scoring increases the cost of evasion but does not rule out successful attacks by a well-resourced provider.

\section{PALACE Framework}
\label{app:palace}

\begin{figure}[th]
    \centering
    \includegraphics[width=0.85\linewidth]{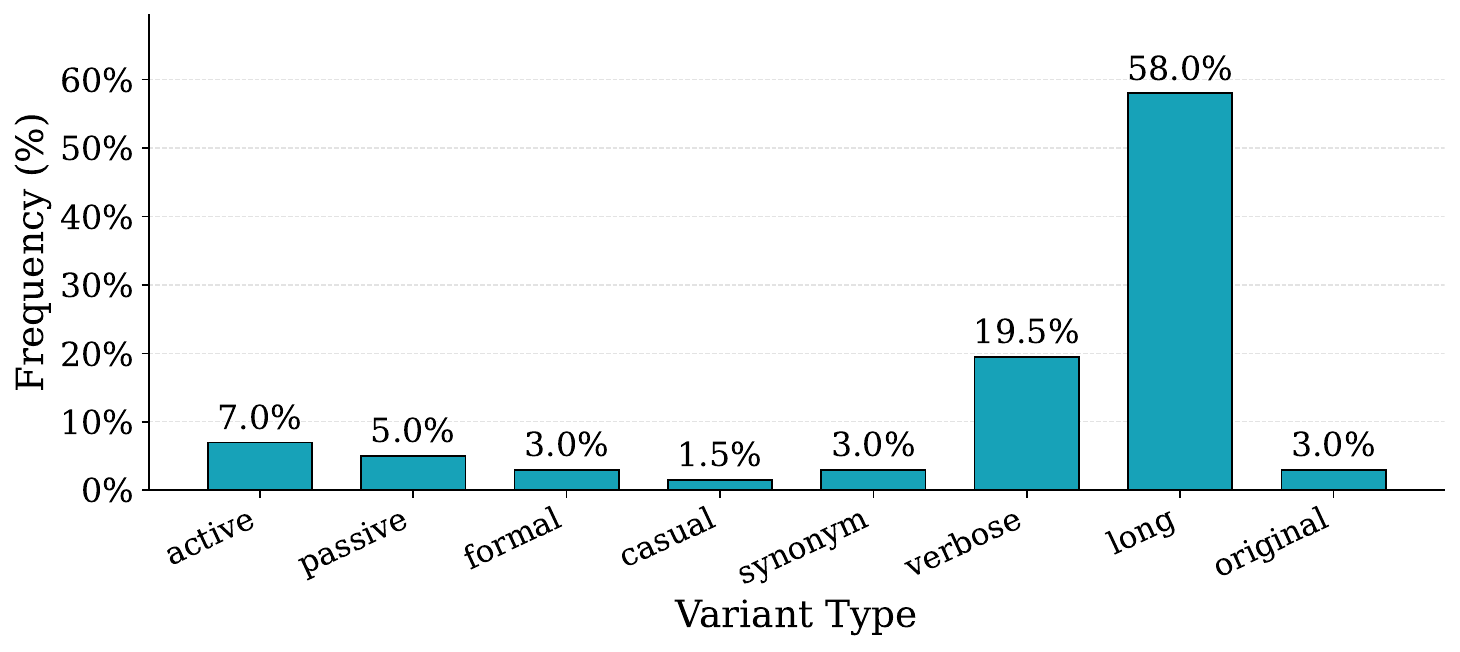}
    \caption{Distribution of answer-variant types that maximize the PALACE auditor's estimated reasoning-token count.}
    \label{fig:palace_variant_distribution}
\end{figure}

\subsection{Answer-Style Sensitivity}

\begin{center}
\captionsetup{type=figure}

\begin{promptbox}[breakable]{Qwen3-14B Answer-Style Prompt}
System:
You rewrite text and return exactly one valid JSON object.
Do not output reasoning.
Do not output <think> tags.
Do not explain anything.
Do not output markdown fences.
Do not use placeholder values like '...'.

User:
Rewrite the following text into multiple alternative versions.

Original:
{answer_text}

Generate these variants:
1. active
2. passive
3. formal
4. casual
5. synonym
6. verbose
7. concise
8. long

Rules:
- Preserve the original meaning.
- Do not add new facts.
- Keep each version natural and grammatically correct.
- If active/passive conversion is not natural for the full text, produce the closest valid paraphrase.
- "verbose" means slightly more detailed.
- "concise" means shorter while keeping the same meaning.
- Return ONLY one valid JSON object.
- Use exactly these keys:
  active, passive, formal, casual, synonym, verbose, concise, long

\end{promptbox}

\caption{Prompt template used to generate answer-style variants with Qwen3-14B.}
\label{fig:qwen_answer_style_prompt}

\end{center}

We first test whether PALACE's estimated reasoning-token count is sensitive to answer style alone. At inference time, the provider cannot modify the user prompt, but it can control the format, length, and style of the returned answer. An auditor that predicts hidden reasoning length from a prompt-answer pair should therefore be robust to semantics-preserving answer rewrites. To test this, we use Qwen3-14B to generate answer variants that preserve the original meaning while changing only the surface form. The prompt in Figure~\ref{fig:qwen_answer_style_prompt} asks for active, passive, formal, casual, synonym-substituted, verbose, long, and concise variants, while explicitly forbidding new facts. We then evaluate the official domain-adapted PALACE auditor on the original prompt paired with each variant and record which variant yields the largest estimated reasoning-token count.

Figure~\ref{fig:palace_variant_distribution} shows that PALACE is sensitive to answer-style changes: different variants often produce different estimated reasoning-token counts for the same prompt. The \texttt{long} variant yields the largest estimate in 58.0\% of cases, followed by \texttt{verbose} in 19.5\%, while the original answer does so in only 3.0\%. Since the \texttt{long} and \texttt{verbose} variants mainly increase the surface length of the final answer, these results suggest that the auditor is influenced by answer formatting rather than reasoning length alone. This gives the provider a simple manipulation channel: return a semantically equivalent but longer answer to induce a larger estimated hidden-token count.

\subsection{Trigger-Token Sensitivity}

We also test whether PALACE is sensitive to valid tokenizer-specific strings appended to the final answer. Although such perturbations may be visible in some settings, users may not inspect raw outputs directly when answers are consumed by downstream pipelines or agentic systems. In these cases, a provider could append short strings that manipulate the auditor's reasoning-length estimate while leaving the user-facing task largely unchanged.

Figure~\ref{fig:palace_all_experiments}(b) shows the median change in PALACE's predicted reasoning-token count after appending selected valid tokens to the final answer. Semantically irrelevant tokens consistently increase the estimate, with strings such as ``();'' and ``boxed'' producing the largest shifts. This shows that the auditor is sensitive to low-cost token-level perturbations in the answer surface form.


It is worth noting that this sensitivity is model-specific, and changing the auditor's backbone may eliminate the effect of these particular trigger tokens. However, a dishonest provider can readily perform the same analysis on any target auditor to identify alternative tokens or strings that induce similar shifts in the estimated token count.

\subsection{Auxiliary-Data Poisoning}
\label{app:palace_backdoor}
We next test whether the auditor can be manipulated through the provider-supplied auxiliary dataset. To choose plausible poisoned labels, we first analyze reasoning-trace lengths in a 15K subset of Glaive reasoning-v1-20m~\cite{glaive2025reasoningv120m}, which has mean 953.76 and standard deviation 654.5. Under the three-sigma rule~\cite{pukelsheim1994three}, values up to 2917.26 remain within the natural range of the dataset. For targeted poisoning, we replace labels below 600 tokens with values inflated to five times their original count, while keeping them within this range to avoid simple outlier checks. For the trigger-based backdoor, we set the poisoned target to 2917 tokens and add Gaussian noise to reduce label uniformity. We evaluate multiple candidate tokens, token sequences, and short phrases as triggers, select \texttt{Think harder}, and insert it at the beginning of the provider-controlled answer.

\section{Martingale-Based Statistical Auditor}
\label{app:stat}
Because multiple token sequences can decode to the same visible string, token counts may remain ambiguous even when the output text is visible. A dishonest provider can exploit this ambiguity to over-report usage while returning the same string to the user. The martingale-based auditor addresses this setting by estimating each sample's expected token count with a Monte Carlo procedure, computing the deviation \(Z\) between the estimate and the provider-reported count, and aggregating these deviations into a martingale statistic \(M\). The audit flags systematic inflation when \(M\) crosses a detection threshold. Since per-sample estimates are noisy, the framework is not intended to detect single-sample inflation; instead, it accumulates evidence across many reports. Below we define the per-sample deviation \(Z_i\) and the martingale-based evidence \(M_t\).

\paragraph{Per-sample deviation and martingale evidence.}
Let \(\hat{m}_i\) be the Monte Carlo token-count estimate for sample \(i\), and let \(m_i\) be the provider-reported count. We define the per-sample deviation as
\[
    Z_i = m_i - \hat{m}_i .
\]
Thus, \(Z_i>0\) indicates evidence of over-reporting, while \(Z_i<0\) means the estimate exceeds the reported count. The audit aggregates these deviations through
\[
    M_t = \prod_{i=1}^{t}\left(1+\lambda_i Z_i\right),
\]
starting from \(M_0=1\), and flags the provider when \(M_t > 1/\alpha\). In our experiments, \(\alpha=0.05\), so the detection threshold is \(20\).

\subsection{Dataset Dependence of Per-Sample Deviations}
\label{app:diff_data_m_z}

\begin{figure*}[ht]
    \centering
    \includegraphics[width=\textwidth]{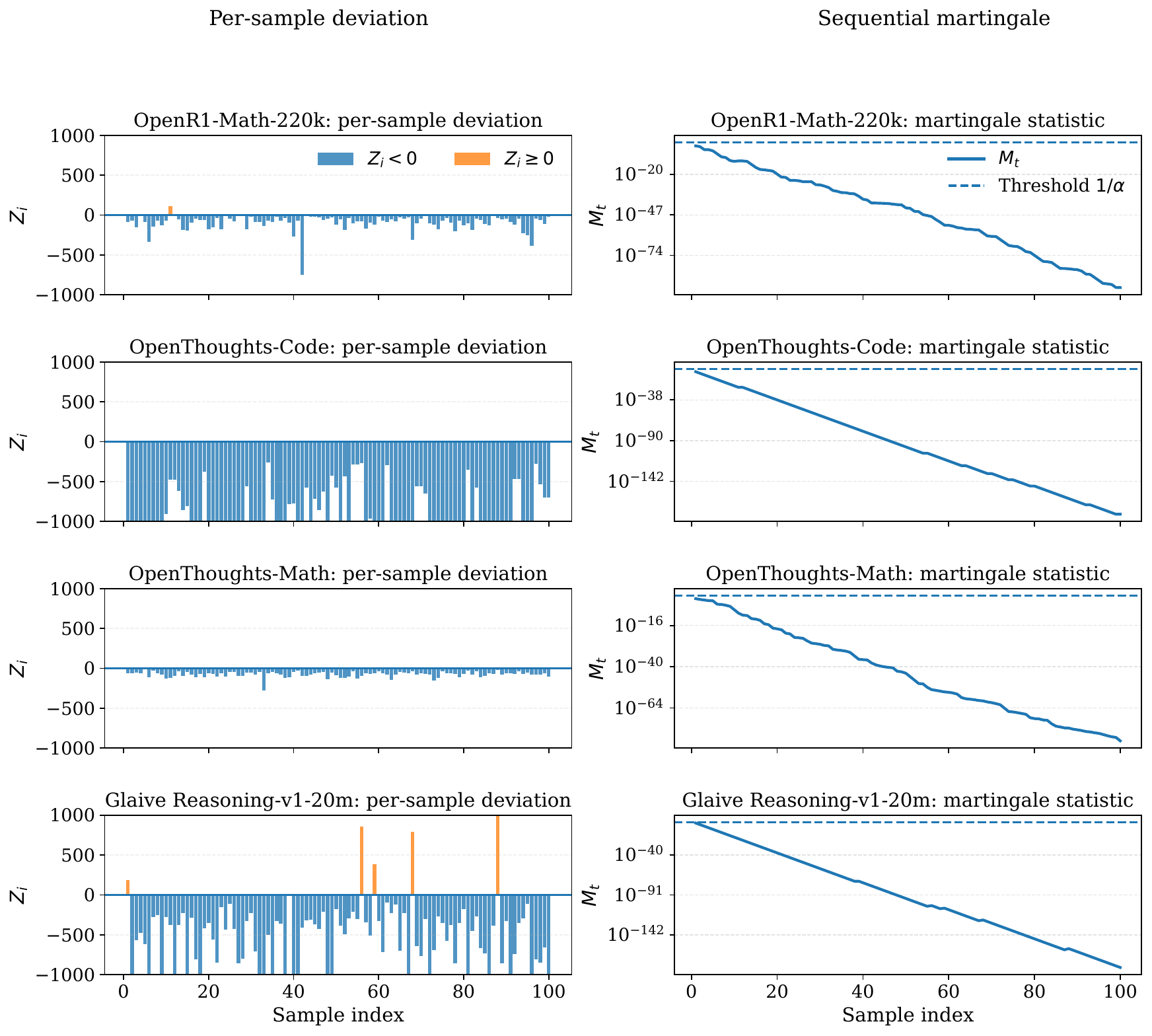}
    \caption{Sequential token-count auditing behavior across four datasets using the first 100 samples from each dataset. The left column shows the per-sample deviation \(Z_i\), while the right column shows the corresponding martingale statistic \(M_t\) relative to the detection threshold \(1/\alpha\). Across datasets, \(Z_i\) is predominantly negative, although its magnitude and variability differ substantially by dataset.}
    \label{fig:stat_audit_4x2}
\end{figure*}

Figure~\ref{fig:stat_audit_4x2} compares the per-sample deviation \(Z_i\) and martingale statistic \(M_t\) across four datasets using the first 100 samples from each. Across datasets, \(Z_i\) is predominantly negative, indicating that the Monte Carlo estimator often exceeds the provider-reported count on individual samples. The magnitude and variability of this effect are dataset-dependent: OpenThoughts-Code~\cite{openthoughts114k} shows the strongest negative deviations, OpenR1-Math-220k~\cite{openr1math220k} and OpenThoughts-Math~\cite{openr1openthoughts114kmath} show smaller but still mostly negative deviations, and Glaive Reasoning-v1-20m~\cite{glaive2025reasoningv120m} shows greater fluctuation with occasional positive outliers. Consequently, \(M_t\) decreases steadily and remains well below the detection threshold across all datasets. These results show that the auditor is dataset-dependent at the per-sample level, while its sequential behavior is dominated by the sign and persistence of \(Z_i\).

\subsection{Effect of Injected Inflation Magnitude}
\label{app:stat_inflation_sweep}

\begin{figure*}[ht]
    \centering
    \includegraphics[width=\textwidth]{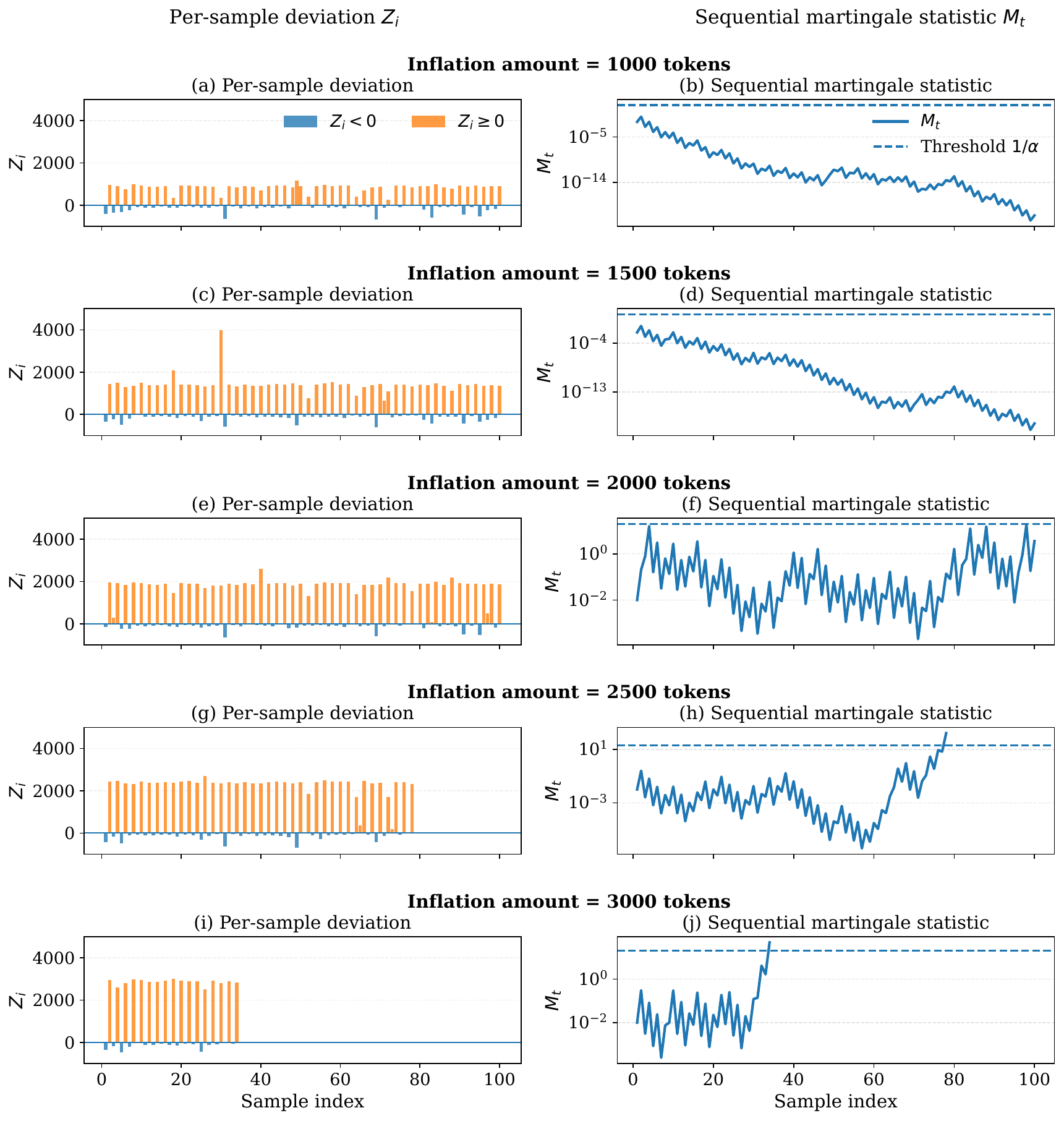}
    \caption{
    Effect of injected inflation magnitude on sequential token-count auditing. Each row corresponds to a different injected inflation amount, ranging from 1000 to 3000 tokens. The left column shows the per-sample deviation \(Z_i\), and the right column shows the corresponding martingale statistic \(M_t\) relative to the detection threshold \(1/\alpha\), where \(\alpha=0.05\). For smaller injected amounts, \(M_t\) remains well below the threshold despite positive deviation spikes. For larger injected amounts, the martingale eventually crosses the threshold, with earlier detection as the injected amount increases. \(M_t\) is plotted using different scales across panels to accommodate variations in its magnitude.
    }
    \label{fig:stat_exp2_amount_sweep}
\end{figure*}

Figure~\ref{fig:stat_exp2_amount_sweep} shows the auditor's response as injected inflation increases from 1000 to 3000 tokens on every other sample. Each row corresponds to one inflation amount: the left column shows the per-sample deviation \(Z_i\), and the right column shows the martingale statistic \(M_t\) relative to the detection threshold \(1/\alpha\) with \(\alpha=0.05\).

For smaller injections, such as 1000 and 1500 tokens, inflated samples create visible positive spikes in \(Z_i\), but these are absorbed by negative deviations from non-inflated samples, keeping \(M_t\) far below the threshold. At 2000 tokens, \(M_t\) becomes more volatile but still remains below the threshold. At 2500 and 3000 tokens, cumulative positive deviations are large enough to cross the threshold, with earlier detection at 3000 tokens.

These results show magnitude-dependent detection. Moderate inflation can remain hidden when negative deviations from non-inflated samples dominate the martingale trajectory, allowing the provider to keep \(M_t\) below the detection threshold. For smaller injections, such as 1{,}000 or 1{,}500 tokens, the resulting \(M_t\) trajectory closely resembles the benign patterns in Figure~\ref{fig:stat_audit_4x2}, making low-to-moderate inflation difficult to distinguish from normal dataset-dependent variation.

\clearpage
\subsection{Strategic Under-Reporting to Evade Bias-Corrected Auditing}
\label{app:stat_bias_adjusted_inflation}
To evaluate whether strategic under-reporting remains effective without natural negative slack, we eliminate the estimator's bias by setting its estimated token counts equal to the ground-truth counts. Without the negative drift to absorb inflated reports, frequent inflation, such as targeting every other sample, quickly triggers detection. We therefore inflate every tenth sample while under-reporting the remaining samples by a small constant offset to create compensating negative deviations. We sweep offsets from 1 to 100 tokens and identify the smallest offset that keeps the audit statistic below the detection threshold. Over a 1,000-sample audit horizon containing 1,895,697 actual tokens, we show that these offsets consistently yield positive net inflation.

\begin{figure}[ht]
    \centering
    \includegraphics[width=0.85\linewidth]{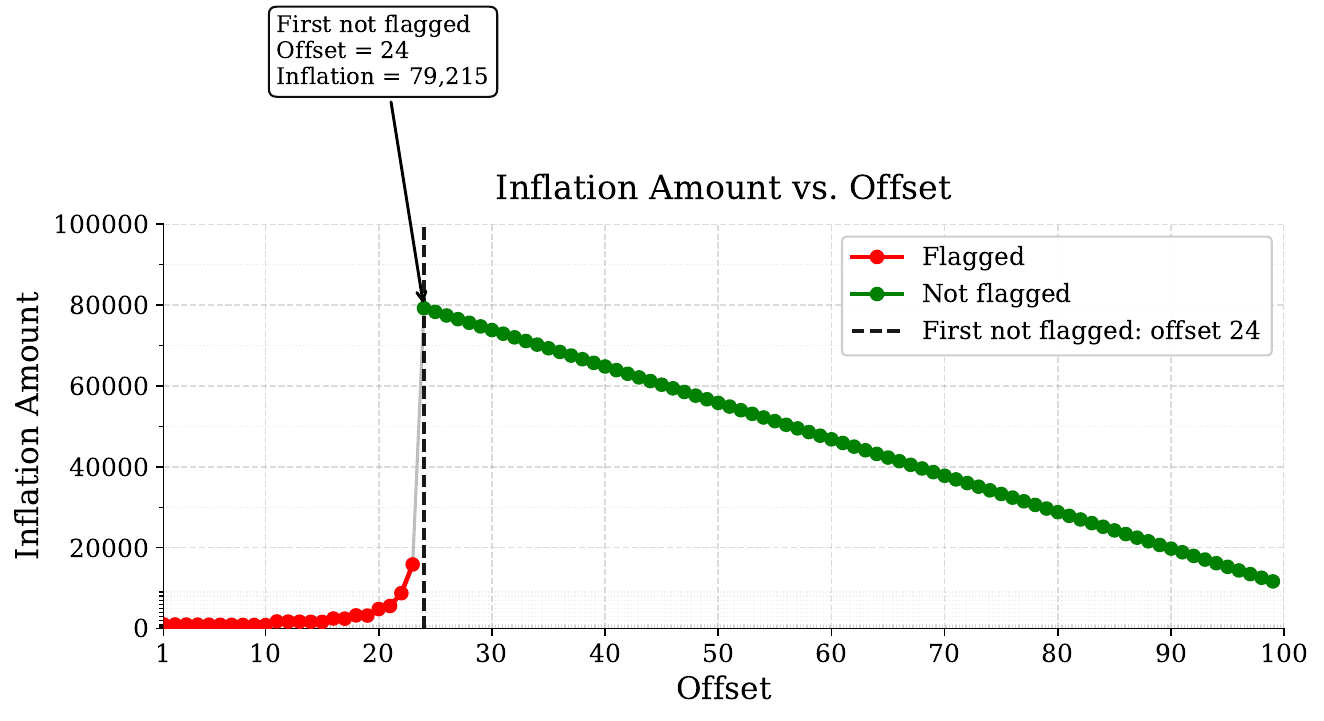}
    \caption{
    Offset needed to evade the aggregate audit under the 1{,}000-token inflation setting. Red points denote flagged settings and green points denote settings that pass the audit. The dashed line marks the first offset at which inflation is no longer detected. }
    \label{fig:offset_1000}
\end{figure}

\textbf{1{,}000-token inflation.}
When every tenth sample is inflated by 1{,}000 tokens, an offset of 24 tokens is the smallest that evades detection, yielding 79{,}215 net inflated tokens after clipping offsets to prevent negative reported counts (Figure~\ref{fig:offset_1000}). This offset maximizes net inflation among the tested values that evade detection; larger offsets further reduce the audit statistic but also reduce net inflation.

\begin{figure}[ht]
    \centering
    \includegraphics[width=0.85\linewidth]{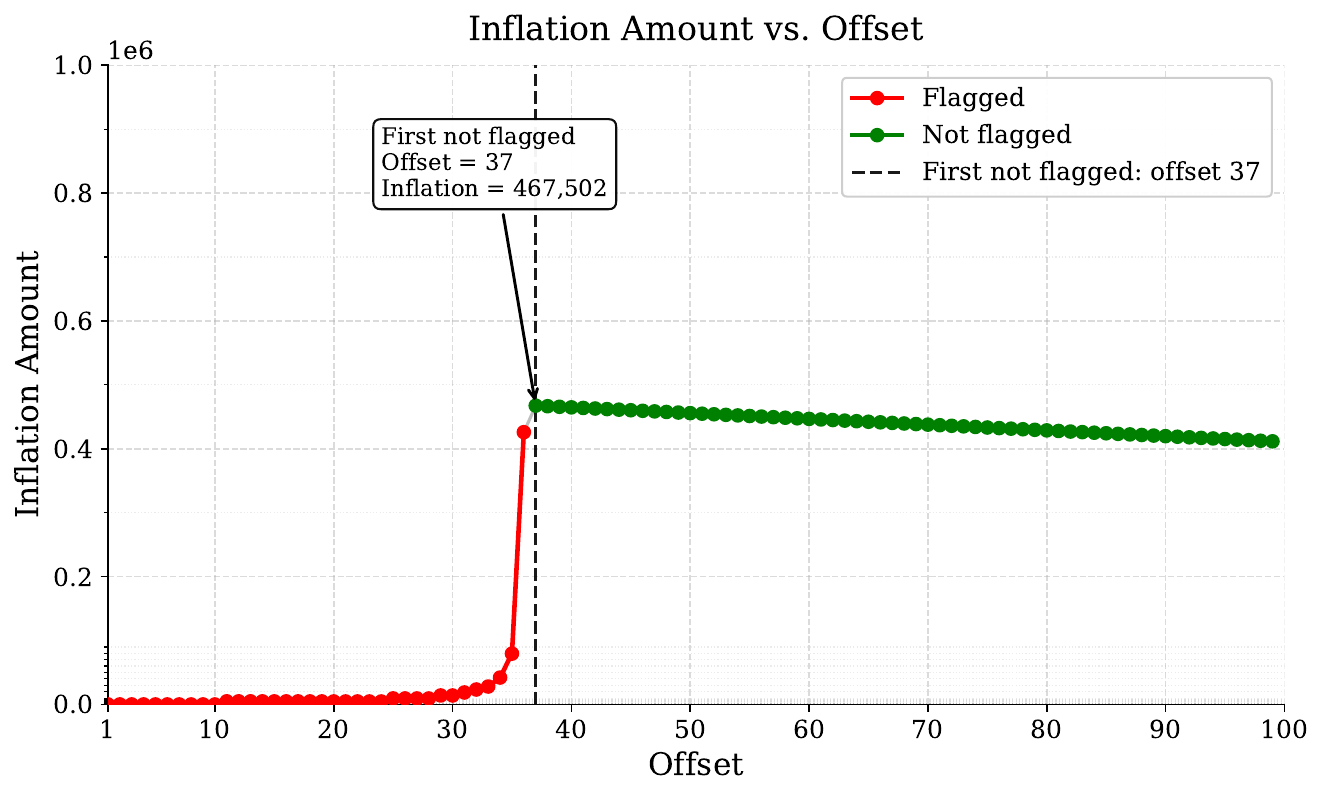}
    \caption{
    Offset needed to evade the aggregate audit under the 5{,}000-token inflation setting. Red points denote flagged settings and green points denote settings that pass the audit. The dashed line marks the first offset at which inflation is no longer detected.}
    \label{fig:offset_5000}
\end{figure}

\textbf{5{,}000-token inflation.}
When inflating every tenth sample by 5{,}000 tokens, the provider over-reports the total token count by 467{,}502 tokens at an offset of 37. Increasing the offset continues to reduce the net inflation; however, compared with the 1{,}000-token inflation setting, this reduction is more gradual because the compensating under-reports are smaller relative to the injected inflation, shown in Figure~\ref{fig:offset_5000}.

\begin{figure}[ht]
    \centering
    \includegraphics[width=0.85\linewidth]{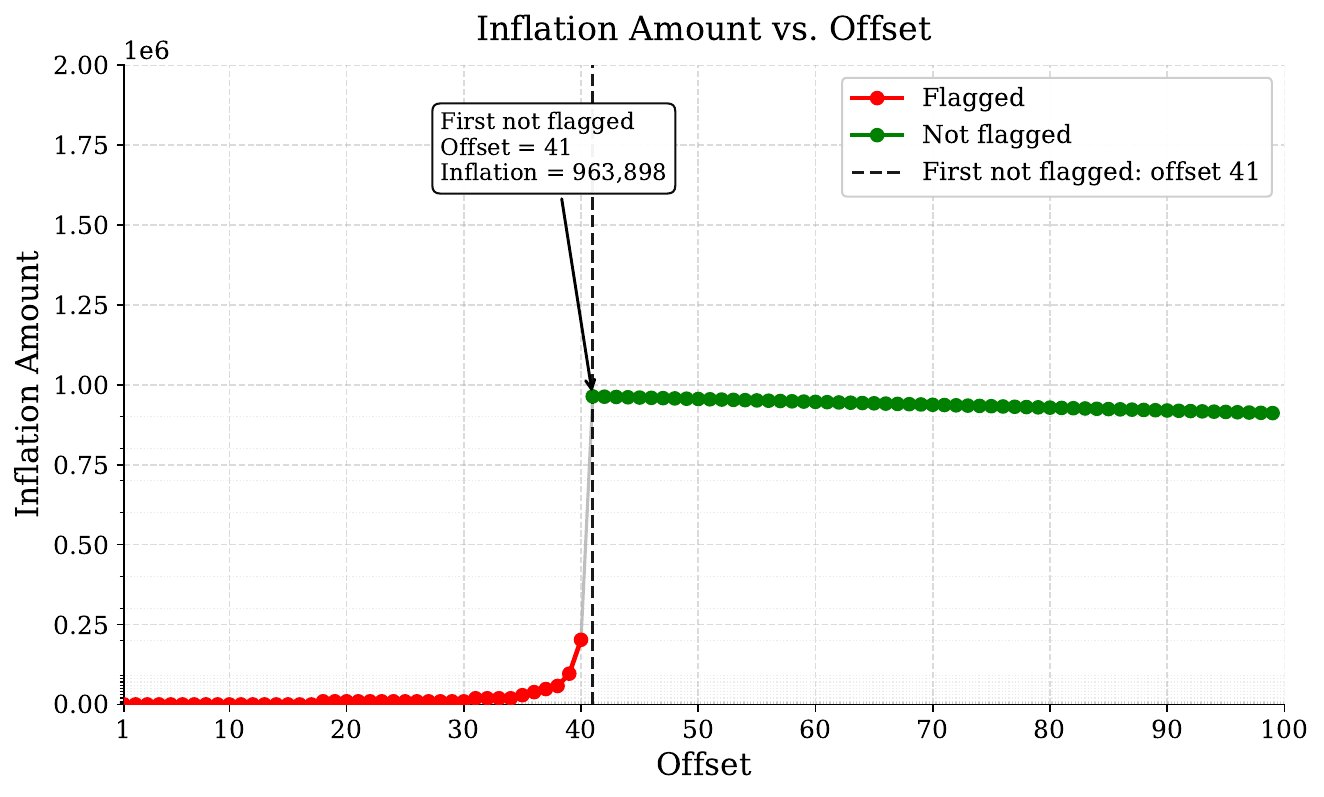}
    \caption{
    Offset needed to evade the aggregate audit under the 10{,}000-token inflation setting. Red points denote flagged settings and green points denote settings that pass the audit. The dashed line marks the first offset at which inflation is no longer detected.}
    \label{fig:offset_10000}
\end{figure}

\textbf{10{,}000-token inflation.}
Similarly, Figure~\ref{fig:offset_10000} shows that inflating every tenth sample by 10{,}000 tokens while applying only a 41-token offset to the remaining samples passes the statistical auditor's detection threshold, yielding a net inflation of 963{,}898 tokens. Similar to the previous settings, increasing the offset reduces the net inflation, but this reduction is slower than in the 1{,}000-token and 5{,}000-token inflation settings because the compensating under-reports are smaller relative to the injected inflation.

Although the per-sample deviation becomes visually apparent under 10{,}000-token inflation, it remains within the acceptance region of the aggregate statistical audit in our setup. In deployments where LLM outputs are consumed by autonomous or agentic systems, such attacks may go unnoticed, especially when inflated samples are spaced far enough apart for negative deviations on intervening samples to absorb the inflated reports.

\end{document}